\documentclass[usenatbib,fleqn]{mnras} 
\usepackage{epsfig}
\usepackage{times}
\usepackage{fixltx2e} 
\usepackage{amsmath,amssymb} 
\usepackage[normalem]{ulem}
  
\newcommand{\Msol}{{\,\rm M}_\odot} \newcommand{\Mpc} {{\,\rm Mpc}}
\newcommand{\kpc} {{\,\rm kpc}} \newcommand{\Gyr} {{\,\rm Gyr}}
\newcommand{\pc} {{\,\rm pc}} 
\newcommand{\kms}{{\,\rm {km\,s^{-1}} }}

\def\Gyr{\,{\rm Gyr}}

\newcommand{\msunyr}{{\,\rm M}_\odot{\rm yr}^{-1}}

\newcommand{\cc}{{\,\rm {cm^{-3}}}}
 
\newcommand{\kmsec}{{\,\rm {km\,s^{-1}} }}

\newcommand{\HI}{H${\rm\scriptstyle I}$ }

\newcommand{\vg}{{\small VINTERGATAN}}
\newcommand{\vgs}{{\small VINTERGATAN }}

\newcommand{\afe}{$[\alpha/{\rm Fe}]$ }
\newcommand{\afet}{$[\alpha/{\rm Fe}]$}
\newcommand{\feh}{$[{\rm Fe/H}]$ }
\newcommand{\feht}{$[{\rm Fe/H}]$}
\newcommand{\Zsun}{{\,\rm Z}_\odot}
\newcommand{\erg}{{\,\rm erg}}

\defcitealias{Renaud2020}{Paper II}
\defcitealias{Renaud2020b}{Paper III}

\title[Structure of a simulated Milky Way-mass galaxy] {VINTERGATAN I: The origins of chemically, kinematically and structurally distinct discs in a simulated Milky Way-mass galaxy}
\author[Oscar Agertz et al.] 
{Oscar Agertz$^1$\thanks{\tt oscar.agertz@astro.lu.se},
Florent Renaud$^1$, Sofia Feltzing$^1$, 
Justin~I.~Read$^{2}$, Nils~Ryde$^1$, \newauthor Eric~P.~Andersson$^{1}$, Martin~P.~ Rey$^{1}$, Thomas~Bensby$^{1}$ and Diane~K.~Feuillet$^{1}$\\
$^1$ Lund Observatory, Department of Astronomy and Theoretical Physics, Lund University, Box 43, SE-221 00 Lund, Sweden\\
$^2$ Department of Physics, University of Surrey, Guildford, GU2 7XH, UK
}
              
\date{\today}
\begin{document}
\maketitle

\begin{abstract}
Spectroscopic surveys of the Milky Way's stars have revealed spatial, chemical and kinematical structures that encode its history. In this work, we study their origins using a cosmological zoom simulation, {\small VINTERGATAN}, of a Milky Way-mass disc galaxy. We find that in connection to the last major merger at $z\sim 1.5$, cosmological accretion leads to the rapid formation of an outer, metal-poor, low-$[\alpha/{\rm Fe}]$ gas disc around the inner, metal-rich galaxy containing the old high-$[\alpha/{\rm Fe}]$ stars. This event leads to a bimodality in $[\alpha/{\rm Fe}]$ over a range of $[{\rm Fe/H}]$. A detailed analysis of how the galaxy evolves since $z\sim 1$ is presented. We demonstrate the way in which inside-out growth shapes the radial surface density and metallicity profile and how radial migration preferentially relocates stars from the inner to the outer disc. Secular disc heating is found to give rise to increasing velocity dispersions and scaleheights with stellar age, which together with disc flaring explains several trends observed in the Milky Way, including shallower radial $[{\rm Fe/H}]$-profiles above the midplane. We show how the galaxy formation scenario imprints non-trivial mappings between structural associations (i.e. thick and thin discs), velocity dispersions, $\alpha$-enhancements, and ages of stars, e.g. the most metal-poor stars in the low-$[\alpha/{\rm Fe}]$ sequence are found to have a scaleheight comparable to old high-$[\alpha/{\rm Fe}]$ stars. Finally, we illustrate how at low spatial resolution, comparable to the thickness of the galaxy, the proposed pathway to distinct sequences in $[\alpha/{\rm Fe}] $-$[{\rm Fe/H}]$ cannot be captured.

\end{abstract}


\section{Introduction}
\label{sect:intro}
Understanding how galaxies form and evolve is a central theme in modern astrophysics. The Milky Way, being the one galaxy that we can study in exquisite detail, provides a fundamental testbed for theories of galaxy formation and evolution \citep[][]{Freeman2002}. Central to the discussion of the Milky Way's origins is the concept of distinct thin and thick stellar disc components \citep[for a review, see][]{RixBovy2013}. A structural dichotomy was first discovered by \citet{GilmoreReid1983} who used star counts towards the South Galactic Pole to
demonstrate that the vertical stellar distribution could not be fit with a single exponential profile. Rather, two exponentials with different scaleheights were required. Such geometrically separated discs have subsequently been observed in other galaxies in the local Universe \citep[e.g.][]{DalcantonBernstein2002}. Spectroscopic studies of Solar neighborhood stars have established that the Milky Way disc also features \emph{chemically} distinct components \citep[][]{Fuhrmann1998,Feltzing2003,Bensby2003,Adibekyan2013,Bensby2014}, specifically between the abundance ratio of $\alpha$-elements relative to iron (\afet) over a wide range of metallicities (\feht). Today this chemical bimodality is observed throughout the Galactic disc \citep[][]{Hayden2015}. 
 
The stellar population with high \afe could have formed rapidly in the early stages of the Milky Way's past when it was compact and gas rich. The proto-galactic gas was then enriched primarily with $\alpha$-elements synthesised and promptly released by core collapse supernovae (SNe). The population of stars in the low-\afe sequence is on the other hand younger, and could form from gas in a more extended disc enriched by type Ia SNe (SNIa) that occur over long timescales \citep[billions of years since the Galaxy started to form,][]{MatteucciGreggio1986}. This general picture is supported by observations where the high-\afe disc population is found to be thicker \citep[][]{Reddy2003}, more radially compact \citep[][]{Bensby2011} and older \citep[$\sim 8-12\Gyr$,][]{Bensby2014,Feuillet2019} than the low-\afe component (which features stars with ages $\lesssim 8\Gyr$). In addition, old $\alpha$-enhanced stars in the Solar neighbourhood are observed to be kinematically hotter \citep[e.g.][]{Holmberg2009}, with a higher mean velocity dispersion ($\gtrsim 40 \kmsec$) than the $\alpha$-poor stars \citep[$\lesssim 20\kmsec$,][]{Lee2011_segue,Hayden2019}.

However, the notion of distinct discs in the Milky Way is still debated. By studying stars with similar elemental abundances, so-called mono abundance populations, \citet{Bovy2012} argued that there is no clear separation between structurally thin and thick discs, but rather a smooth transition (see also \citealt{Bovy2016}). In essence, there is not always a well defined mapping between the structural associations (thick or thin disc), $\alpha$-enhancements, or ages of stars \citep[see also][]{Mackereth2017,Minchev2017}. 

These complex structural and chemo-dynamical trends may uniquely encode how the Galaxy formed and evolved \citep[][]{Freeman2002,BlandHawthorn2019}, which is why it is important to develop a theoretical framework for their origins. Several mechanisms for creating structurally thick and kinematically hot discs exist in the literature, including vertical disc heating by satellite encounters \citep[][]{Quinn1993,Kazantzidis2009}, accretion of satellite stars \citep[][]{Abadi03a,Read2008}, star formation in turbulent gas rich discs \citep[][]{Bournaud09}, heating by giant molecular clouds \citep[e.g.][]{Aumer2016}, and secular formation by radial migration of kinematically hot stars from the inner to the outer disc \citep[][]{SellwoodBinney2002,Loebman2011}. 

A number of theories for how chemically distinct discs form have been put forward over the past decades. We highlight two models from the literature that are both able to reproduce abundance gradients in the Milky Way disc as well as broadly explain observed abundance trends in the Solar neighbourhood. The first was pioneered by \citet[][]{Chiappini1997} (for variations and refinements, see e.g. \citealt{Haywood2013}, \citealt{Spitoni2019}) and is often referred to as the `two-infall model'.  In this model an initial phase of star formation, rapid enough to be unpolluted by SNIa, results in the formation of the high-\afe sequence. Star formation is then assumed to proceed inefficiently with the remaining gas becoming enriched by SNIa, reducing its \afet. A second period of low metallicity gas infall subsequently lowers the metallicity of the interstellar medium. This leads to the buildup of a low-\afe population of stars that overlaps with the high-\afe sequence in terms of \feht. The second model is by \citet{SchonrichBinney2009} and postulates that the two chemically distinct components represent the equilibrium star formation conditions in different parts of the disc. Radial migration of stars via cold torquing, also known as `churning', by a bar and spiral waves \citep[][]{SellwoodBinney2002, Roskar08, Minchev2013} then allows for the populations to spatially overlap in the Solar neighbourhood.

Galaxy chemical evolution models, such as the one by \citet[][]{Chiappini1997}, are useful tools to in a computationally efficient way probe the roles of various physical processes. While informative, they are phenomenological by construction with many degrees of freedom and unknown parameters such as gas infall rates and star formation depletion timescales, which limit their predictive power \citep[for a recent overview, see][]{Andrews2017}. As a complement to such models, cosmological hydrodynamical simulations of galaxy formation come with less simplifying assumptions, although with a higher computational cost and the added complexity of directly modelling processes such as star formation and stellar feedback robustly \citep[for a review, see][]{NaabOstriker2017}. Furthermore, low numerical resolution has made cosmological simulations unsuited for studying the internal structure of galaxies, leading to predominantly thick, kinematically hot discs \citep[][]{House2011}. This hurdle has only recently been overcome, with several authors reporting on thin, kinematically cold disc components \citep[e.g.][]{Bird2013,Martig2014,AgertzKravtsov2015,Grand2017,Hopkins2018,Buck2020a, Bird2020}.

Generic predictions of cosmological simulations of Milky Way-mass spiral galaxies are an inside-out formation scenario of stellar discs \citep[][]{Abadi03a,Agertz2011,Brook2012thinthick,Stinson2013,GarrisonKimmel2018} due to the steady buildup of angular momentum from large scale tidal torques \citep[][]{Peebles69,FallEfstathiou80,Pichon2011} coupled to a non-destructive contribution of accreting gas to the disc's angular momentum reservoir \citep[][]{Sales2012,Kretschmer2020}. Furthermore, cosmological simulations tend to reproduce the observed `upside-down' disc formation scenario, with old stars residing in thicker discs compared to young stars \citep[]{Bird2013}. This is due to disc thickening from early epochs of (gas rich) mergers \citep[e.g.][]{Brook2004}, in addition to the secular heating processes discussed above.

Despite this progress, the situation is less clear for chemically distinct discs, with the mere existence of a chemical bimodality being difficult to reproduce in a cosmological context. \citet{Mackereth2018} used the large volume {\small EAGLE} simulation and concluded that the scarcity of simulated galaxies exhibiting distinct \afe sequences indicates that the Milky Way cannot be representative of the broader population of disc galaxies. Zoom simulations of Milky Way-like galaxies, that can reach higher numerical resolution, have also been unable to recreate this Milky Way feature \citep[e.g.][]{Ma2017}, raising the question as to whether it is a rare feature of disc galaxies. 

In contrast, a number of authors have recently reported on higher fractions of \afe bimodalities in simulations of disc galaxies. A range of explanations for their origins have been put forward, including gas-rich mergers \citep[][]{Brook2012thinthick,Grand2018,Buck2020b} and rapid star formation in high redshift clumps \citep[][]{Clarke2019}, with varying conclusions regarding the importance of secular processes such as radial migration \citep[][]{Minchev2013}. Clearly, a consensus regarding the formation channels of chemically distinct galactic disc components, and their connection to structural and kinematical thin and thick discs has not yet been reached. New generations of cosmological simulations are required to interpret observations and advance our theoretical understanding.

In this first paper in a series, we make use of a new high resolution cosmological simulation to identify the physical mechanisms that lead to the formation of chemical, kinematical and spatial thick and thin discs in a Milky Way-mass galaxy\footnote{Movies are available at \\\url{http://www.astro.lu.se/~florent/vintergatan.php}}. In a companion paper, \citet{Renaud2020} (hereafter \citetalias{Renaud2020}), we extend on the work presented here and identify the contributions of the in situ and accreted material as well as role of galaxy interactions and mergers. In \citet{Renaud2020b} (hereafter  \citetalias{Renaud2020b}), we explore the role played by the assembly of an extended, outer gaseous disc at $z>1$ in forming the most metal-poor stars in the simulated galaxy.

This paper is organized as follows. In Section \ref{sect:method} we present the numerical method, including our choice of galaxy formation physics and simulation setup. In Section \ref{sect:generalprop}, we outline general properties of our simulated galaxy, followed by an overview and formation scenario of its chemical structure in Sections \ref{sect:chembimod} and \ref{sect:whathappened}. In Sections \ref{sec:discgrowth} and \ref{sect:vertical}, we present a detailed analysis of how the internal structure evolves over the past 8 billion years. We connect the chemical, kinematical and spatial structures using mono-abundance populations in Section \ref{sect:MAP}. Finally we discuss and conclude our results in Sections \ref{sect:discussion} and \ref{sect:conclusion}, respectively.

\section{Method}
\label{sect:method}

\subsection{Simulation setup}
\label{sec:setup}
We base our analysis on a cosmological hydrodynamic+$N$-body zoom-in simulation of a Milky Way-mass galaxy carried out with the adaptive mesh refinement code {\small RAMSES} \citep{teyssier02}. 

We first performed a dark matter-only simulation with $512^3$ particles in a periodic box with size $85 \Mpc$ assuming a flat $\Lambda$-cold dark matter cosmology with \mbox{$H_0 = 70.2 \kms \Mpc^{-1}$}, $\Omega_{\rm m} = 0.272$, $\Omega_\Lambda = 0.728$, and \mbox{$\Omega_{\rm b}= 0.045$}. The initial conditions were generated with the {\small MUSIC} code \citep{music2011}. We note that these are the same initial conditions as the `m12i' halo from \citet{Hopkins2014} and \citet{Wetzel2016}, drawn from the volume used in the AGORA galaxy formation comparison project \citep{agora, agora2}. At $z=0$, a dark matter halo with $R_{200,{\rm m}} = 334 \kpc$ (radius of a sphere with a density 200 times the mean cosmic matter density) and virial mass $M_{200,{\rm m}} = 1.3\times 10^{12} \Msol$ was identified. The halo experiences its last major merger at a lookback time of $\sim 9\Gyr$, in agreement with what we know about the Milky Way's history \citep[][]{Ruchti2015}. Particles inside $3 R_{200,{\rm m}}$ at $z=0$ were then traced back to $z=100$ where the numerical resolution in the Lagrangian volume was increased \citep[for details, see][]{music2011}, resulting in dark matter particles with masses of $3.5\times 10^4\Msol$ and gas mass resolution of $7070\Msol$. 

Mesh refinement is based on a pseudo-Lagrangian approach, where a cell is split if its baryonic mass (gas and stars) exceeds 8 times the initial gas mass resolution. In addition, a cell is allowed to refine if it contains more than 8 dark matter particles. This allows the local force softening to closely match the local mean inter-particle separation, which suppresses discreteness effects \citep[e.g.][]{Romeo08}. The maximum refinement level is set to allow for a mean constant physical resolution of $\sim 20 \pc$ in the dense interstellar medium (ISM).

For the hydrodynamics, we use the HLLC Riemann solver
\citep{Toro1994} and the MinMod slope limiter to construct gas
variables at cell interfaces from their cell-centred values. To close
the relation between gas pressure and internal energy, we use an ideal gas equation of state with an adiabatic index $\gamma=5/3$. 

With the high resolution region embedded in the full (lower resolution) cosmological box, the simulation was evolved with hydrodynamics and galaxy formation physics (see Section~\ref{sect:galformphysics}) to $z=0.17$ where it was stopped due to the high numerical cost. Motivated by the slow, gradual secular evolution of the galaxy at late times, we use the last simulation snapshot as a proxy for present-day conditions.

\subsection{Galaxy formation physics}
\label{sect:galformphysics}
The adopted star formation and feedback physics is presented in \citet{Agertz2013} and \citet{AgertzKravtsov2015, agertzkravtsov2016}. Briefly, star formation is treated as a Poisson process, sampled using $10^4\Msol$ star particles, occurring on a cell-by-cell basis according to the star formation law, 
\begin{equation}
	\dot{\rho}_{\star}= \epsilon_{\rm ff}\frac{ \rho_{\rm g}}{t_{\rm ff}} \quad {\mbox{for}} \quad  \rho>\rho_{\rm SF}\quad {\rm and} \quad T_{\rm gas}<T_{\rm SF}, 
	\label{eq:schmidtH2}	
\end{equation}
Here $\dot{\rho}_{\star}$ is the star formation rate density, $\rho_{\rm g}$ the gas density, $t_{\rm ff}=\sqrt{3\pi/32G\rho_{\rm g}}$ is the local free-fall time, $\rho_{\rm SF}=100~{\rm cm}^{-3}$ is the star formation threshold, $T_{\rm SF}=100~{\rm K}$ is the maximum allowed temperature of star forming gas, and $\epsilon_{\rm ff}$ is the local star formation efficiency per free-fall time of gas in the cell. The efficiency is computed following the relation from \citet{Padoan2012}, derived from simulations of star formation in magnetized supersonic turbulence\footnote{$\epsilon_{\rm ff}=0.5\exp(-1.6 t_{\rm ff}/t_{\rm dyn})$, where $t_{\rm dyn}=L/2\sigma$ is the dynamical time and $\sigma$ is the local velocity dispersion. The latter is computed using neighbouring gas cells over a region of size $L=$ 3 grid cells per spatial dimension.}. This favours star formation in unstable gas, with a majority of stars in the simulation found to form at very high densities, $\rho\gtrsim 10^4~{\rm cm}^{-3}$, rather than close to the adopted threshold. The resulting mean efficiency is $\epsilon_{\rm ff}\sim 1$ per cent, with local values varying by over two orders of magnitude, in agreement with observed star formation efficiencies in the Milky Way's giant molecular cloud population \citep[][]{Lee2016}.

Each formed star particle is treated as a single-age stellar population with a \citet{chabrier03} initial mass function. We account for injection of energy, momentum, mass and heavy elements over time from core collapse SN, SNIa and stellar winds and radiation pressure on the surrounding gas. Each mechanism depends on stellar age, mass and gas/stellar metallicity \citep[through the metallicity dependent age-mass relation of][]{Raiteri1996}, calibrated on the stellar evolution code {\small STARBURST99} \citep{Leitherer1999}. The modelled stellar mass loss leads to star particles losing up to 50 per cent of their mass over a Hubble time \citep[see also][]{Leitner2011}.

\begin{figure*}
\centering
\includegraphics[width=0.98\textwidth]{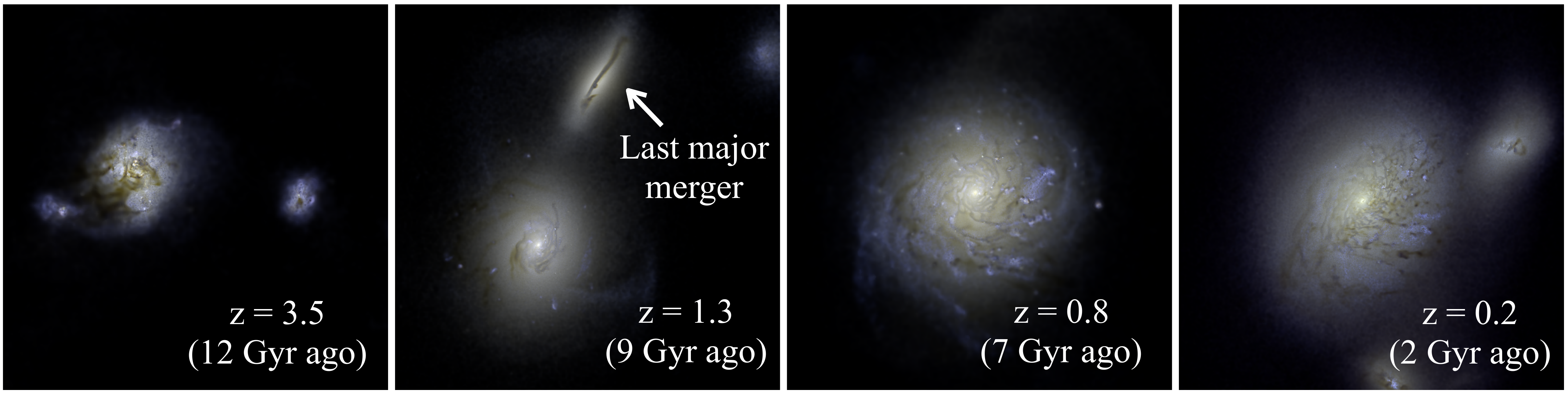}
\caption{Mock HST/ACS and XMM images of the main galaxy at $z=3.5,1.3,0.8$ and 0.2. The RGB composite images are constructed using F814W (red),  F606W (green) and UVW1 (blue) broadband filters. At $z>3$ the gas rich galaxy is irregular due to frequent mergers. At $z\sim 1.3$, the main galaxy (now with a stellar mass $\approx 3\times 10^{10}\Msol$) can be seen to interact with a galaxy 1/3 of its mass. This is the last major merger (indicated in the $z=1.3$ panel), after which the galaxy grows secularly, ending up as an extended disc galaxy.
}
\label{fig:mocks}
\end{figure*}

\begin{figure*}
\centering
\begin{tabular}{cc}
\includegraphics[width=0.44\textwidth]{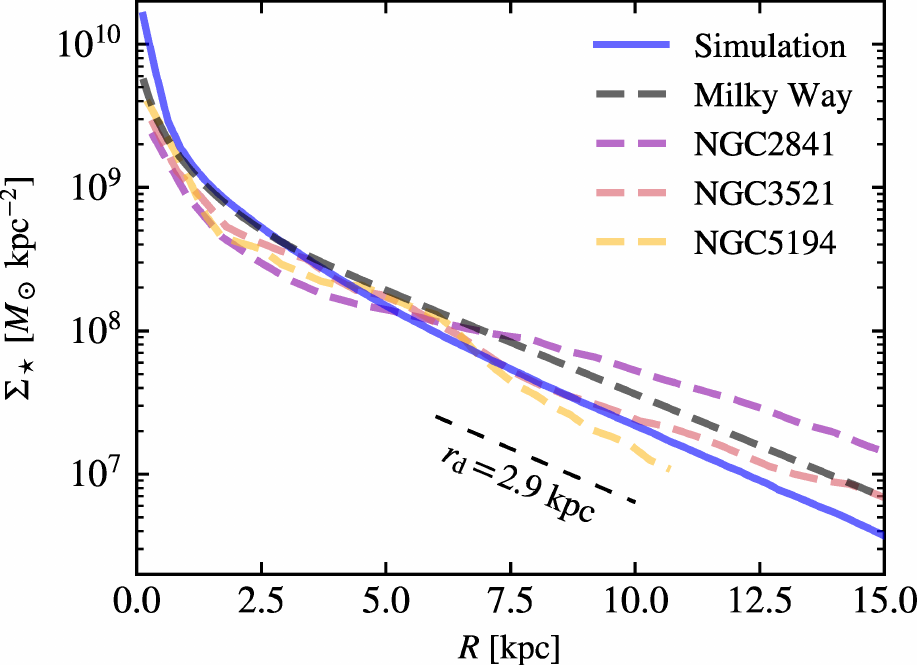} &
\includegraphics[width=0.448\textwidth]{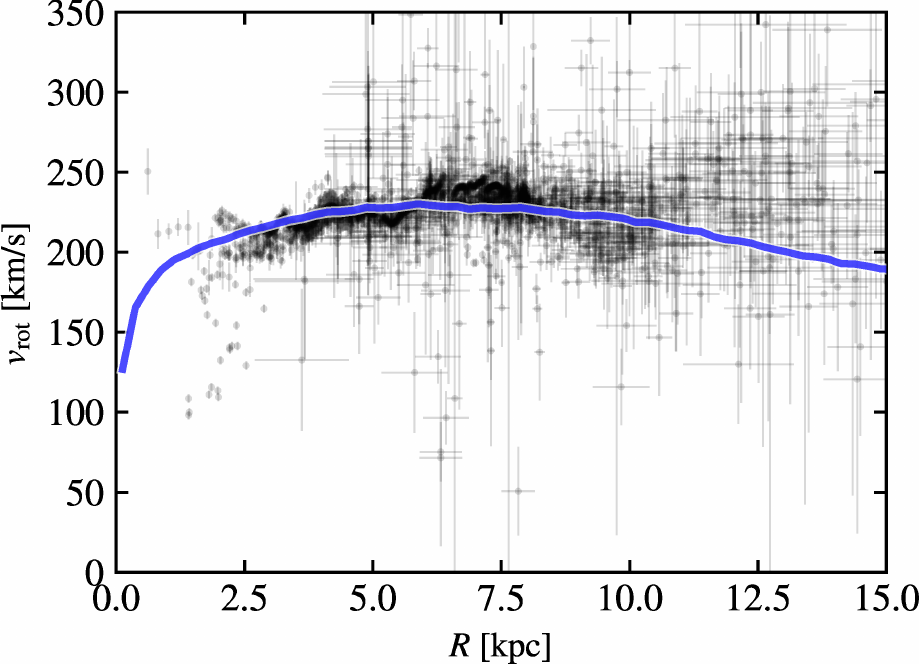}  
\end{tabular}
\caption{Left: Comparison of the surface density profiles of stars in the simulated galaxy, at the final simulation time, along with the profiles of late type galaxies from the THINGS sample \protect\citep{leroy_etal08} in the stellar mass range $M_\star\approx 3-10\times 10^{10}\ \rm M_{\odot}$ and the Milky Way \protect\citep[best-fitting parameters for the thin and thick stellar discs and bulge from][]{McMillan2011}. The exponential scalelength for the simulated galaxy's disc at the current epoch is $r_{\rm d}=2.9~\kpc$. Right: Simulated rotational velocity profile compared to Milky Way data, including the kinematics of gas, stars and masers in a total of 2780 measurements,
by \citet{Pato2017}.}
\label{fig:sigmavrot}
\end{figure*}

\begin{figure}
\centering
\includegraphics[width=0.47\textwidth]{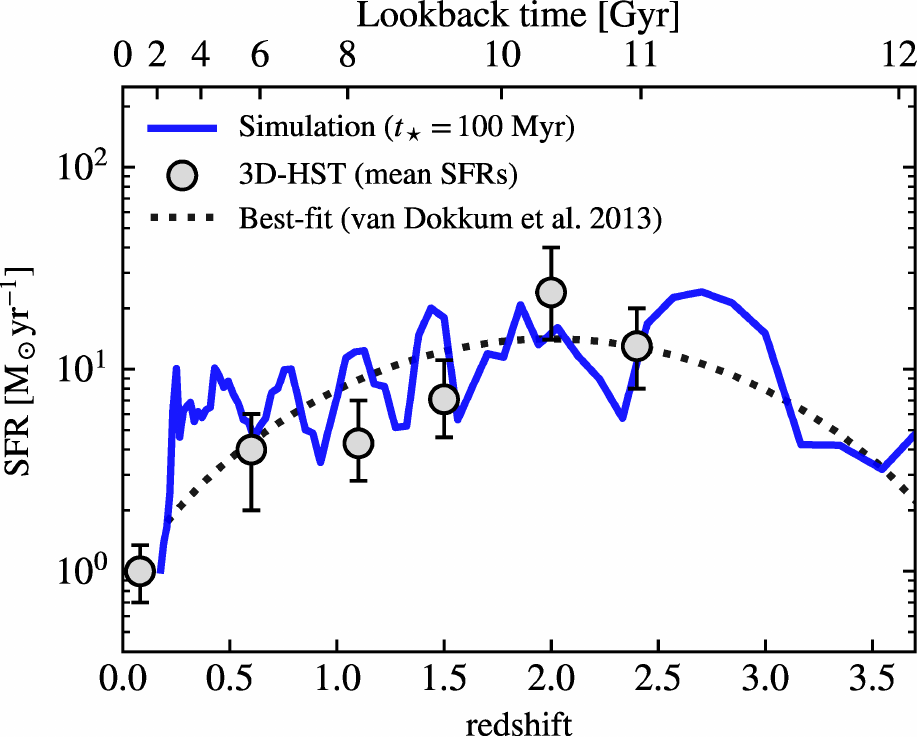} 
\caption{Evolution of the instantaneous star formation rate in the simulated galaxy compared to observations. The dashed line is the best fit relation of the implied $\mathrm{SFR}$ evolution from observed progenitors of star forming galaxies with present-day stellar masses of $\log(M)\approx 10.7$ using the 3D-HST and CANDELS Treasury surveys \citep[see][]{vandokkum2013}. Data points are the mean measured star formation rates of the galaxies in a number of redshift bins, from the 3D-HST v2.1 catalogs \citep[][]{Skelton2014}, adapted from \citet{vandokkum2013}.}
\label{fig:sfh}
\end{figure}

To account for the effect of SN feedback, we adopt the model suggested by \citet[]{KimOstriker2015}. They demonstrated \citep[see also][]{Martizzi2015} that in order to capture the momentum injection from individual SNe, the cooling radius\footnote{The cooling radius in gas with density $n$ and metallicity $Z$ scales as $\approx 30 (n/1\cc)^{-0.43} (Z/Z_\odot + 0.01)^{-0.18} \pc$ for a SN explosion with energy $E_{\rm SN}=10^{51}$ erg \citep[e.g.][]{Cioffi1988, Thornton1998}.} must be captured by at least three grid cells to avoid numerical overcooling. In this work we adopt six grid cells per cooling radius as a minimum requirement for SNe to be considered as resolved. If this criterion is fulfilled, we initialize the explosion in the `energy conserving' phase by injecting $10^{51} \erg$ per SN into the nearest grid cell. When insufficient resolution is available, the explosion is instead initialized in its `momentum conserving' phase, with the momentum built up during the Sedov-Taylor phase injected into cells surrounding the star particle. The adopted relation for the momentum is $4\times 10^5 (E_{\rm SN}/10^{51}\erg)^{16/17} (n/1~{\rm cm}^{-3})^{-2/17} (Z/Z_\odot)^{-0.2} {\Msol \kms}$  \citep[e.g.][]{Blondin1998, KimOstriker2015, Hopkins2018}, where $E_{\rm SN}$ is the total energy injected by SNe in a cell with gas density $n$ and metallicity $Z$ compared to Solar ($Z_\odot=0.02$).

We track iron (Fe) and oxygen (O) abundances separately, with yields taken from \citet{woosley_heger07}. When computing the gas cooling rate, which is a function of total metallicity, we construct a total metal mass following
\begin{equation}
M_{Z}=2.09M_{\rm O}+1.06M_{\rm Fe}
\label{eq:met}
\end{equation}
according to the mixture of alpha (C, N, O, Ne, Mg, Si, S) and iron (Fe, Ni) group elements for the Sun \citep{Asplund2009}. The code accounts for metallicity dependent cooling by using the cooling functions by \citet{sutherlanddopita93} for gas temperatures of $10^{4-8.5}$~K, and rates from \citet{rosenbregman95} for cooling down to lower temperatures. Heating from the ultraviolet background radiation is modelled following \citet{haardtmadau96}, assuming a reionization redshift of $z=8.5$ \citep[implemented following][]{Courty2004}, with gas self-shielding following \citet{AubertTeyssier2010}. Finally, we adopt an initial gas metallicity $Z = 10^{-3} \Zsun$ (accounted for purely in oxygen, see discussion in \citealt{Agertz2020}) in the high-resolution zoom-in region in order to account for enrichment from unresolved population III star formation \citep[see e.g.][]{Agertz09b,Wise2012}. 

\subsection{Abundance ratios}
A focus of our study is the relation between different elemental abundance ratios and how, when, and where these are built up. Specifically, we will trace \afe and \feht, where $\alpha$ stands for `alpha-element(s)'. The $\alpha$-elements relative to iron are understood to track the relative contribution of core collapse SNe and SN\,Ia \citep{Matteucci2001}. In observational studies of stars in the Milky Way, $\alpha$-elements are generally taken to be one or several of the elements O, Mg, Si, Ca, and Ti and an $\alpha$-elemental abundance is then calculated as a mean of some or all of these elemental abundances \citep[see, e.g. ][]{Edvardsson1993}. Sometimes, the overall abundance of all $\alpha$-elements is derived directly in the analysis \citep[e.g.,][]{Hayden2015}.

Ideally, several of these elements should be traced as their formation channels differ \citep[see e.g. discussions in][]{Carlin2018}. However, it is computationally expensive to include a large number of element in a simulation like {\small VINTERGATAN}. We therefore use oxygen as an approximation for all $\alpha$-elements \citep[see also][]{Segers2016,Mackereth2018}. We compute chemical abundances for a star particle as 

\begin{equation}
{\rm [Y/X]}=\log_{10}\left(\frac{f_Y/m_Y}{f_X/m_X}\right)-(\log_{10}\epsilon_{Y,\odot}-\log_{10}\epsilon_{X,\odot})
\end{equation}
where $Y$ and $X$ are the considered elements, $m_Y$ and $m_X$ are their respective atomic masses, and $f_X$ and $f_Y$ are their respective metal mass fractions. Abundances relative to Solar ($\epsilon_{X,\odot}$ and $\epsilon_{Y,\odot}$) are taken from \citet{AndersGrevesse1989}, where Solar mass fractions of O and Fe are 0.0097 and 0.00185 respectively.

\section{Results}
\label{sect:results}
\subsection{General properties and comparison to observations}
\label{sect:generalprop}
We begin by highlighting a number of moments during the simulated galaxy's history using the composite multi-wavelength maps presented in Fig.~\ref{fig:mocks}. At $z>3$ (lookback time $>11$ Gyr) mergers are frequent, leading to an irregular morphology. Over the subsequent billions of years, a compact star forming disc starts to emerge. At a lookback time of $\sim 9\Gyr$, a galaxy with a stellar mass $M_\star\approx 10^{10}\Msol$ interacts with the main progenitor ($M_\star \approx 3\times 10^{10}\Msol$), and the two galaxies coalesce at $z\sim 1.2$ ($\approx 8.5\Gyr$ ago). This is the last major merger (hereafter LMM), and it marks the epoch when a bimodality in \afet-\feh begins to form. As we will demonstrate below, the bimodality arises from cosmological gas accretion at this epoch, and is hence an \emph{indirect} effect of the LMM. At $z<1$, only minor mergers take place, and stars form in an extended secularly growing disc through to the current epoch. The final galaxy is a spiral galaxy with $M_\star=6\times 10^{10}\Msol$ (measured within $20$ kpc) residing in a dark matter halo with virial mass \citep[following the definition by][]{bryan97} $M_{200}=10^{12}\Msol$. As such, the galaxy formation efficiency is $M_\star/M_{200}\approx 6$ per cent, in agreement with results from abundance matching \citep[e.g.][]{Moster2010,Behroozi2013,kravtsov_etal14}. 

Next, we compare the simulation to a number of observed disc galaxy characteristics, including properties of the Milky Way. We view this as a necessary `validation step' to warrant a closer examination of the galaxy's internal structure. The left panel of Fig.~\ref{fig:sigmavrot} shows the stellar surface density profile of the simulated galaxy, together with data from late type galaxies from the THINGS sample \citep{leroy_etal08} in the stellar mass range $3-10\times 10^{10}\ \rm M_{\odot}$, as well as the Milky Way. The latter uses a combination of best-fit parameters for the thin and thick stellar discs and bulge from \protect\citet{McMillan2011}. The simulation produces a surface density profile in excellent agreement with local late type galaxies, and features the same exponential scalelength\footnote{Defined via the exponential surface density profile $\Sigma(R)=\Sigma_0\exp{(-R/r_{\rm d})}$, with $\Sigma_0$ being the central density.} ($r_{\rm d}=2.9\kpc$) as found by \citet{McMillan2011} for the Milky Way's thin disc. 

We note that unlike the Milky Way, the simulated galaxy does not feature a bar which can hamper an in-depth structural comparison of the central kiloparsecs. The final galaxy's neutral gas fraction (atomic and molecular hydrogen) is $12$ per cent, which is compatible with the Milky Way \citep[][]{ferriere01} as well as spiral galaxies of similar stellar mass \citep[][]{Catinella2010,Dutton2011}. 

The right panel of Fig.~\ref{fig:sigmavrot} shows the rotational velocity ($v_{\rm rot}$) as a function of galactocentric radius ($R$) compared to Milky Way data from the extensive literature compilation of $v_{\rm rot}$ by \cite{Pato2017}\footnote{The Milky Way data assumes that the distance from the Sun to the Galactic centre is $R_0 = 8$ kpc, the Suns rotational velocity $v_{\rm 0,rot} = 230\kms$ and peculiar Solar motions $(U, V, W )_\odot = (11.10, 12.24, 7.25) \kms$.}. The simulation's rotational velocity profile was computed at the final simulation time, considering the tangential velocity of stars formed in the last $2\Gyr$. Near identical results are recovered when considering the motion of atomic gas. The rotation curve rises to $v_{\rm rot}\approx 230\kms$ at $R=6-7$ kpc and broadly matches the shape and normalisation of the Milky Way's rotation curve, although large observational uncertainties exist beyond $R\gtrsim 10$ kpc. 

Finally, we turn to the star formation history ($\mathrm{SFH}$) of the galaxy shown in Fig.~\ref{fig:sfh}. The star formation rate ($\mathrm{SFR}$) is computed considering only the mass in stars ($M_{\rm \star,young}$) formed within the last $t_\star=100$ Myr of each simulation snapshot, with $\mathrm{SFR}\equiv M_{\rm \star,young}/t_\star$. This allows for a direct comparison to the mean observed star formation rate of galaxies with stellar masses $\log(M_\star)\approx 10.7$ up to redshifts $z\sim 2.5$ from the 3D-HST v2.1 catalogs \citep[][]{Skelton2014} adapted from \citet{vandokkum2013}, with the dashed line showing the best-fit relation. The simulated $\mathrm{SFH}$ features a number of peaks and valleys, indicating epochs of starburst activity triggered by mergers and galaxy interactions, episodically reaching\footnote{The precise peak SFRs depend on the adopted $t_\star$ (observationally, the considered star formation rate tracer).} a $\mathrm{SFR}\sim 20-30\msunyr$ at $z>1$. The overall normalisation and shape of the SFH matches observations, furthering the notion that the simulated galaxy can be viewed as representative of late-type spiral galaxies.

The galaxy's size, luminosity and rotational velocity make it compatible with observed scaling relations such as the Tully-Fisher and size-luminosity relations \citep[][]{Courteau2007}. Furthermore, its specific angular momentum content is compatible with observed late-type spiral galaxies \citep[$j_{\star}\gtrsim 1000 \kpc \kms$,][]{Fall2013}. Finally, \citet{Rhodin2019} demonstrated, using this particular simulation, that its circum-galactic medium (CGM) features strong \HI absorbers, observed as damped Ly-$\alpha$ system in background quasar spectra, at impact parameters and metallicities in agreement with observations.

We emphasize that this simulation does not aim at reproducing the Milky Way star-by-star. Nonetheless, our simulation, which we call \vg\footnote{Swedish for the Milky Way, translating literally to `The Winter Street'.}, matches a range of observed characteristics of galaxies with luminosities and masses similar to the Milky Way. This motivates an in-depth study of how its internal chemical, kinematical and structural properties came to be, which we turn to in the next sections. In future work we will study how changes to the cosmological merger history \citep[see e.g.][]{Pontzen2017,Rey2019,Rey2020} affects such observables. 

\begin{figure}
\centering
\begin{tabular}{cc}
\includegraphics[width=0.44\textwidth]{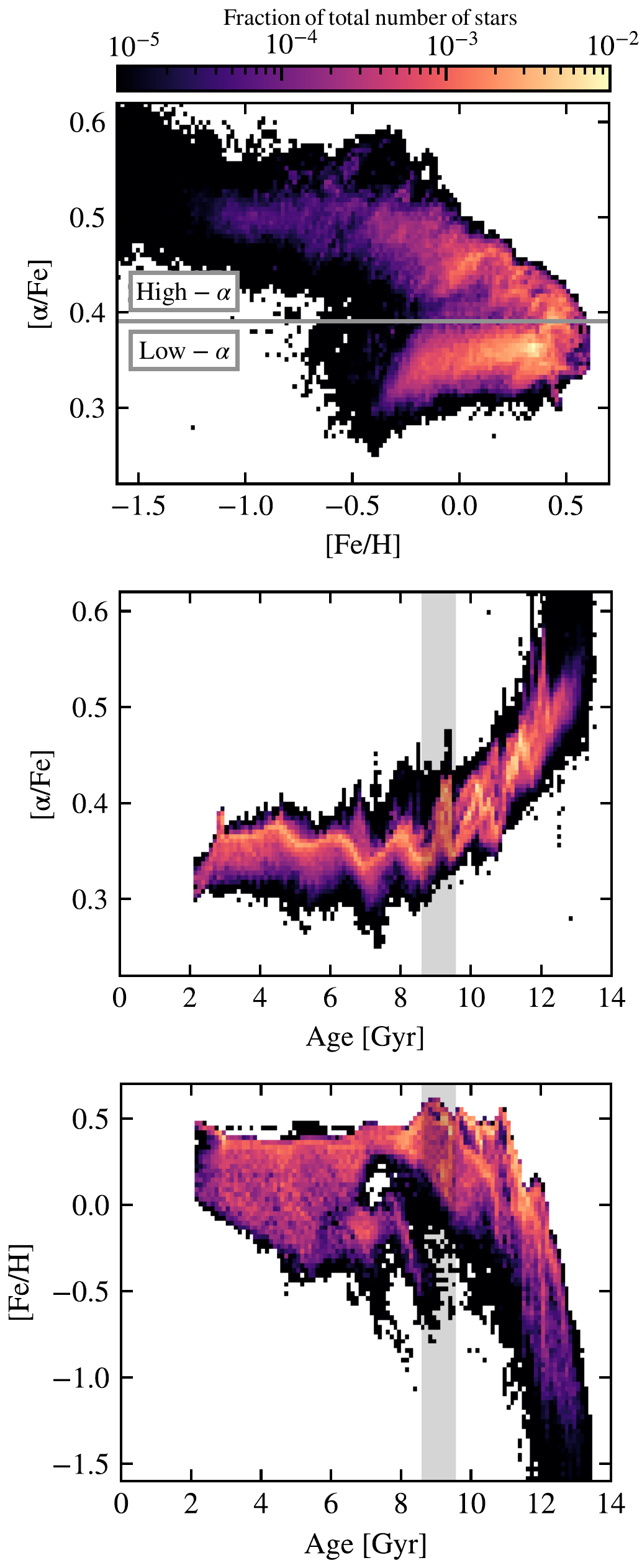} &
\end{tabular}
\caption{Chemical properties of the disc at its final instance, including all stars at $R<20~\kpc$ and $|z|<3\kpc$. Top: \afe vs. \feht. The distribution features a distinct bimodality in \afe across a range of \feht. The horisontal gray solid line indicates the separation between the low-and high-\afe sequences. Middle: \afe vs. age of stars. \afe decreases over time as SNIa progressively enrich the ISM until the epoch of the last major merger, indicated by a gray band, after which stars form in the `low-\afe' sequence. Bottom: \feh vs. stellar age. Before the last major merger (gray band), the galaxy's metallicity reaches \feh$>0$ in the inner disc. At the epoch of the last major merger, gas inflows lead to the formation of an outer disc, visible as a low-\feh feature.
}
\label{fig:chembimodal}
\end{figure}

\begin{figure*}
\centering
\includegraphics[width=0.92\textwidth]{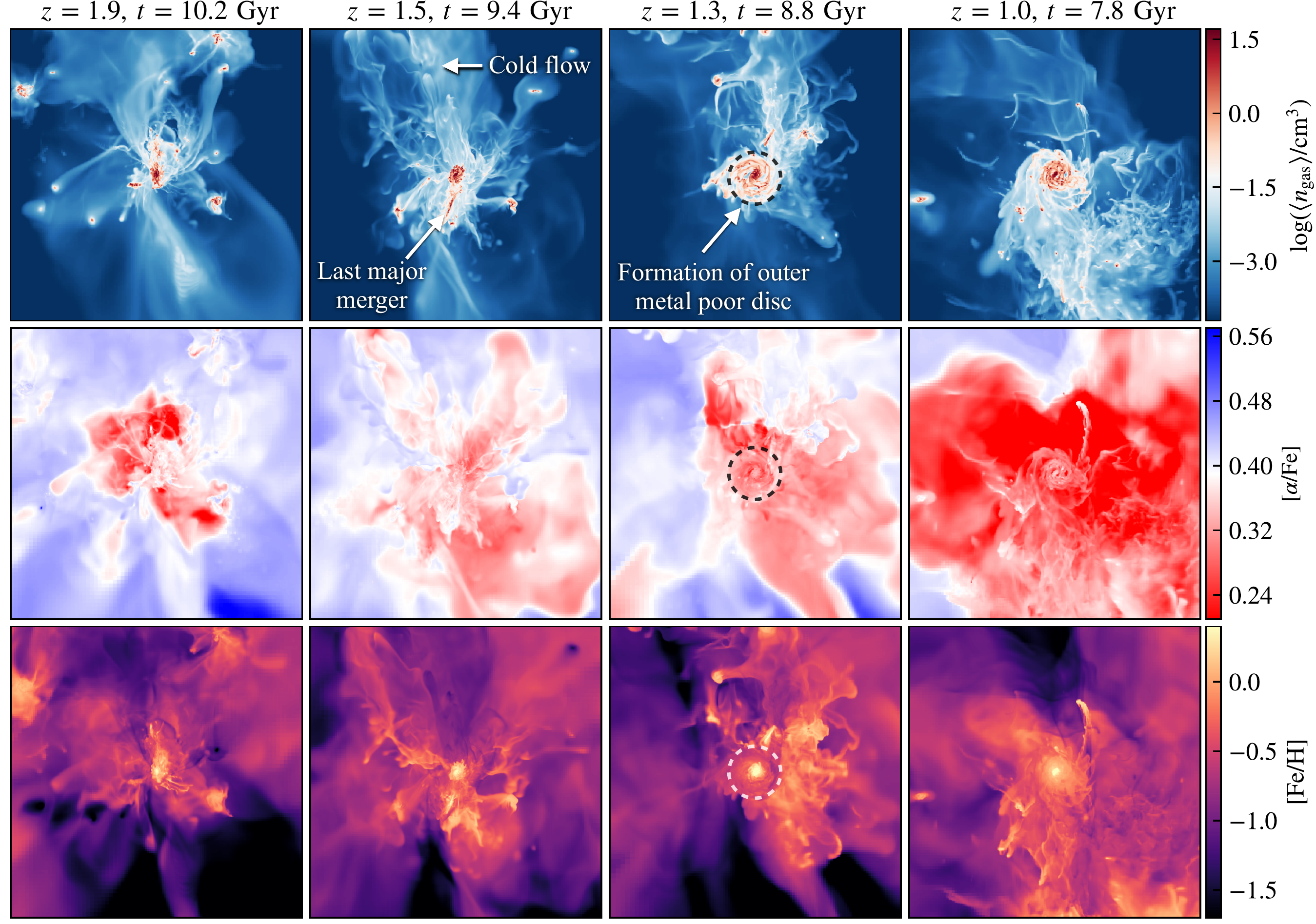} 
\caption{Large scale (140 kpc across) gaseous environment around the galaxy at, from left to right, $z=1.9, 1.5, 1.3$ and $1.0$. From top to bottom, the rows show density-weighted average gas densities, \afe and \feh along the line of sight. The \afet-maps have their (diverging) colourmaps centered around \afet$ =0.39$, the approximate divide between the high-and low-\afe stellar sequences.}
\label{fig:largescale}
\end{figure*}

\begin{figure}
\centering
\includegraphics[width=0.44\textwidth]{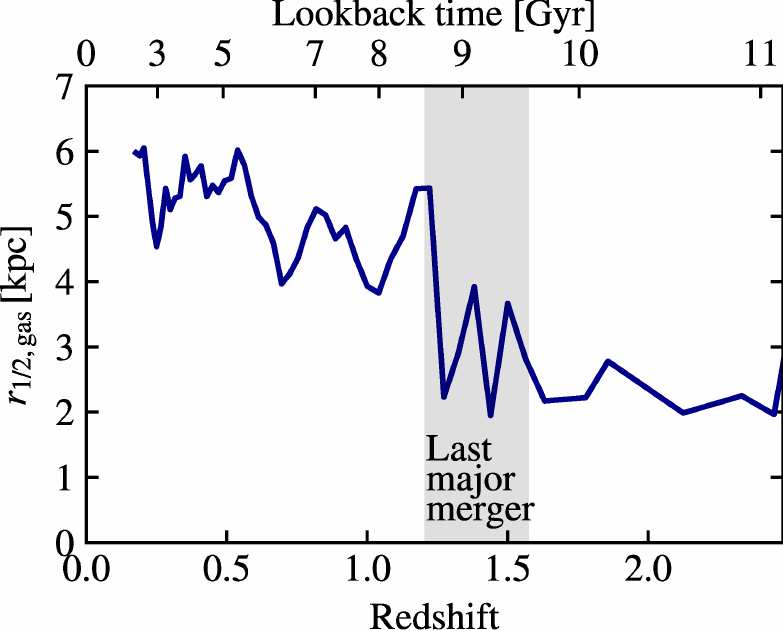} 
\caption{Half-mass radius of the cold star forming ISM ($T<10^{4}~{\rm K}$, $n>1~{\rm cm^{-3}}$).
}
\label{fig:growth}
\end{figure}

\subsection{A chemically bimodal galaxy}
\label{sect:chembimod}
At the last instance of the simulation, we select all stars at galactocentric radii $<20\kpc$ and heights $<3\kpc$ from the midplane for analysis. Stellar ages and lookback times are henceforth computed in relation to $z=0$. The top panel in Fig.~\ref{fig:chembimodal} shows the \afet-\feh plane, and we begin by noting that this galaxy is $\sim 0.3$ dex (i.e. roughly a factor of two) more enhanced in $\alpha$-elements as compared to the Milky Way. This does not only stem from the specific accretion and outflow history of the galaxy, but also from details of the adopted IMF, stellar yields, and rate of SNIa explosions \citep[see e.g.][]{Marinacci2014b,Naiman2018,Philcox2018}. We discuss this topic further in Section~\ref{sect:numuncert}, but leave an in-depth investigation of these ingredients for future work. 

In qualitative agreement with the Milky Way, the distribution is bimodal in \afet. Two distinct sequences can be seen, separated by 0.1-0.2 dex in \afe around \afe$\sim0.39$ for $-0.7 \lesssim$\feh$\lesssim 0.5$. At high \feht, the two sequences connect. The chemical plane is structured, which stems from merger events and starbursts triggered by interactions (in \citetalias{Renaud2020}, we link each notable feature to its physical cause). This burstiness is also visible in the \afet-age relation in the middle panel, where the epoch of the LMM (9 Gyr ago, $z\sim 1.2-1.6$) is indicated with a grey band. On average, \afe decreases over time as SNIa enrich the ISM/CGM. After the epoch of the LMM all stars form in the low-\afe sequence, with a narrow range in \afet.

In contrast, stars form with a range of metallicities at all instances in the galaxy's past, as shown in the \feht-age relation in the bottom panel of Fig.~\ref{fig:chembimodal}. A number of features are worth highlighting in this panel. Before the LMM, the galaxy's \feh has gradually grown to supersolar values. Subsequently, a low-\feh feature forms, extending to \feh$\sim -0.7$. This feature entails \emph{simultaneous} low and high-\feh star formation in a narrow range of \afe (see middle panel of Fig.~\ref{fig:chembimodal}). Furthermore, it marks the onset of a chemically bimodal galaxy, as stars formed with the same low \feh a few billion years prior are $\alpha$-enhanced. Following this event, stars form, for a few billion years, in two well separated populations in terms of \feht, with few stars around \feh$\sim0$. We next demonstrate how these chemical properties relate to the way in which the galaxy's disc assembled.

\begin{figure}
\centering
\includegraphics[width=0.47\textwidth]{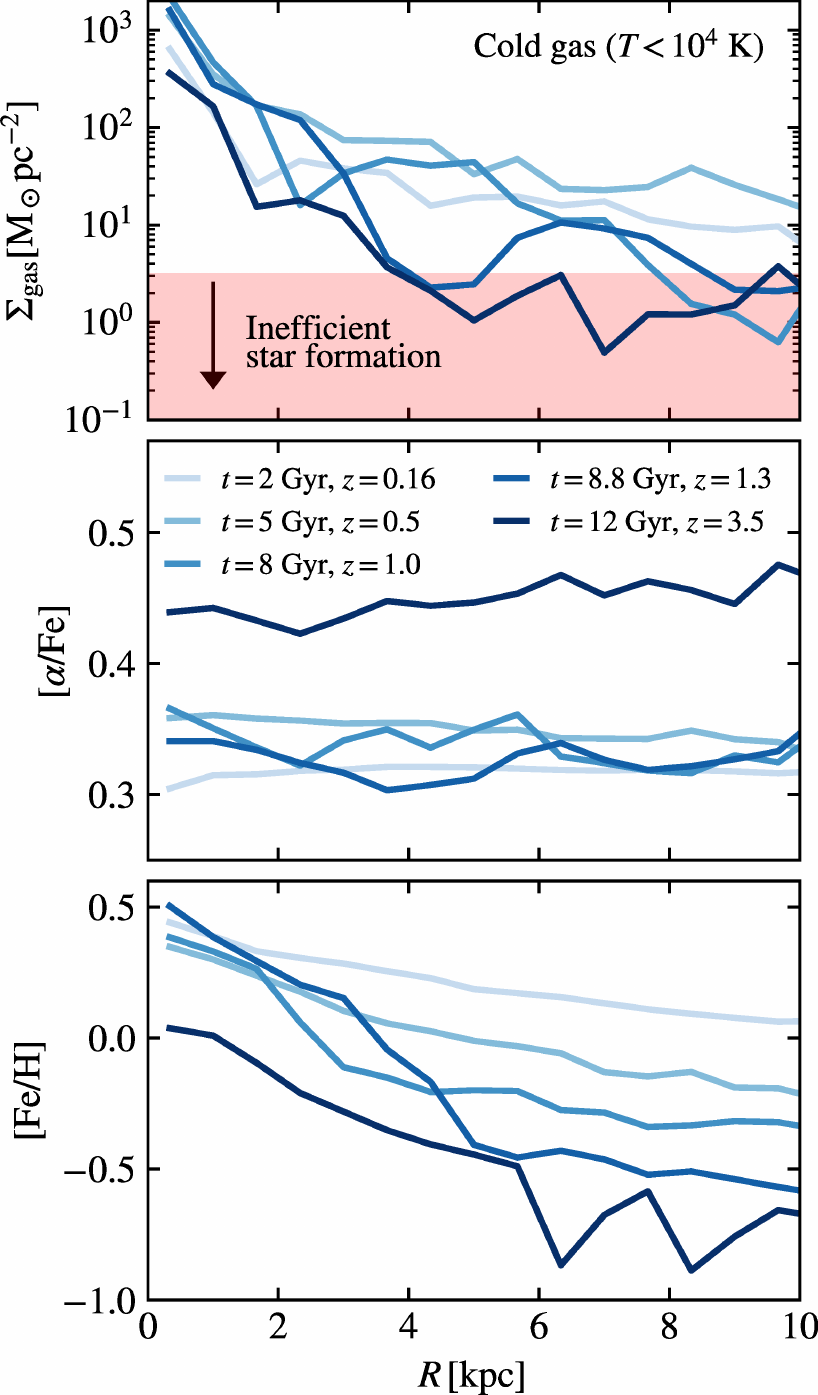} 
\caption{From top to bottom: evolution of surface density, radial \afet-and \feht-profiles of the cold atomic ($T<10^4$K) ISM, the gas phase representing the star-forming disc. The red region in the top panel indicates inefficient star formation, see the main text.}
\label{fig:discprofiles}
\end{figure}

\subsection{How does the bimodality form?}
\label{sect:whathappened}
Fig.~\ref{fig:largescale} shows the large-scale (140 kpc across) gaseous environment around \vgs at $z=1, 1.3, 1.5$ and $1.9$. From top to bottom, the panels show density-weighted averages along the line of sight\footnote{Each pixel value is computed as $\int \rho(l)\cdot A(l) {\rm d}l/\int \rho(l) {\rm d}l$, where $l$ is the position along the line of sight, $\rho$ is the gas density and $A$ is the quantity of interest.} for gas density, gas \afe and gas \feht. 

At $z>1.5$, the galaxy is compact with a metal-rich (\feh$\gtrsim 0$) ISM. The CGM is $\alpha$-enhanced (shown in blue in the middle panels) due to core collapse SNe-enriched outflows at early times, with feedback driven bubbles of SNIa-enriched low-\afe gas (shown in red). Vigorous feedback-driven outflows and the destructive nature of mergers have not yet allowed for an extended disc to form \citep[for a more recent analysis on disc growth vs. merger rate, see ][]{Dekel2019}. Large scale tidal torques continuously increase the angular momentum content in the gas that accretes onto the growing halo \citep[][]{Pichon2011}, a process that favours the formation of extended discs at late cosmic times \citep[e.g.][]{Sales2012, Ubler2014, agertzkravtsov2016}.

At $z\sim 1.5$, cold ($T\lesssim 10^4\,$K), metal-poor (\feh$\lesssim -1$) large scale gaseous flows, together with gas lost from satellites interacting with the gaseous halo, reach the inner parts of the dark matter halo. Some of this gas dilutes the disc, leading to a lowering of the highest attainable stellar \feht, as seen in the bottom panel in Fig.~\ref{fig:chembimodal}. The infalling gas is rich in angular momentum, leading to the formation of an extended disc over the next few 100 Myr, as seen in the $z=1.3$ panels in Fig.~\ref{fig:largescale}. This event is rapid, as quantified in Fig.~\ref{fig:growth} which shows how the half-mass radius of the cold ISM increases from $r_{\rm 1/2,gas}\sim 2-3$ kpc at $z\gtrsim 1.5$ to $\sim 4-6$ kpc at $z\lesssim1.3$: a two-fold increase in size in just $500$ Myr. In \citetalias{Renaud2020b}, we present an in-depth analysis of this event, as well as the physical processes that trigger the onset of the low-\afe sequence from its metal-poor end. This mechanism is central in the assembly of the bimodality and the extended disc, which we briefly outline next.

The newly formed structure is an outer, metal-poor (\feh$\lesssim -0.5$, see $z=1.3$ panel in Fig.~\ref{fig:largescale}) detached gas disc, with an angular momentum vector that does not coincide with that of the inner disc's. Furthermore, the disc is low in \afe due to the low-\afe nature of the filamentary gas flows and the gas accreting from the CGM (shown in red in the middle panels). This results in the formation of the most metal-poor stars in the galaxy's low-\afe sequence. By $z=1$, the outer disc has become massive enough to contain large star forming clumps, and is surrounded by a low-\afe CGM out to $\sim 50-100\kpc$.

\begin{figure}
\centering
\includegraphics[width=0.48\textwidth]{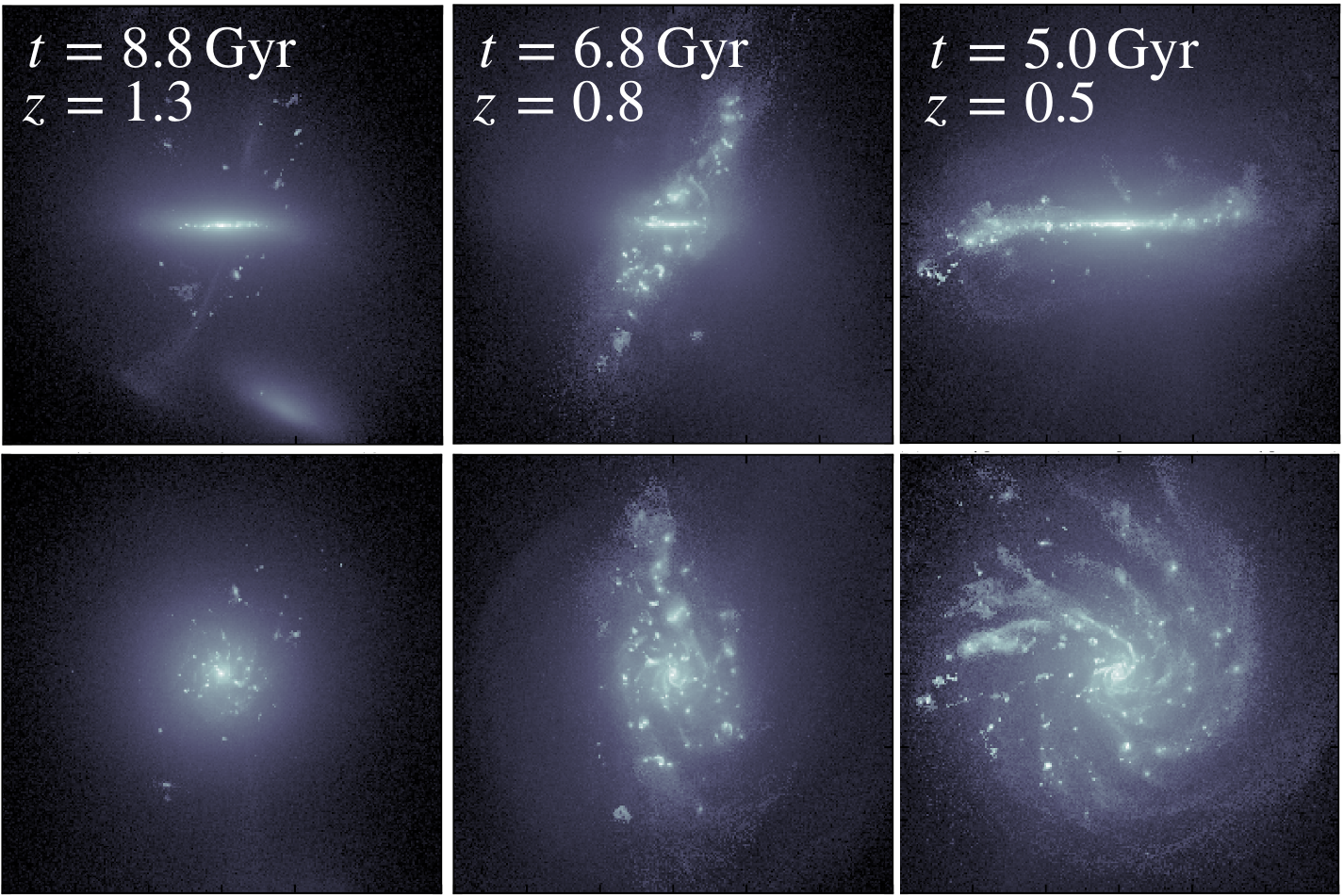} 
\caption{The evolution of the stellar disc seen edge-on (top) and face-on (bottom). Each panel is $24\times24\kpc^2$ in size and shows the bolometric luminosity of the stars. The outer metal-poor disc forms almost orthogonally to the inner disc at $z\sim 1.3-1.5$, with gravitational torques causing them to closely align within $\sim 3-4\Gyr$.}
\label{fig:discs2}
\end{figure}

\begin{figure*}
\begin{center}
\includegraphics[width=0.83\textwidth]{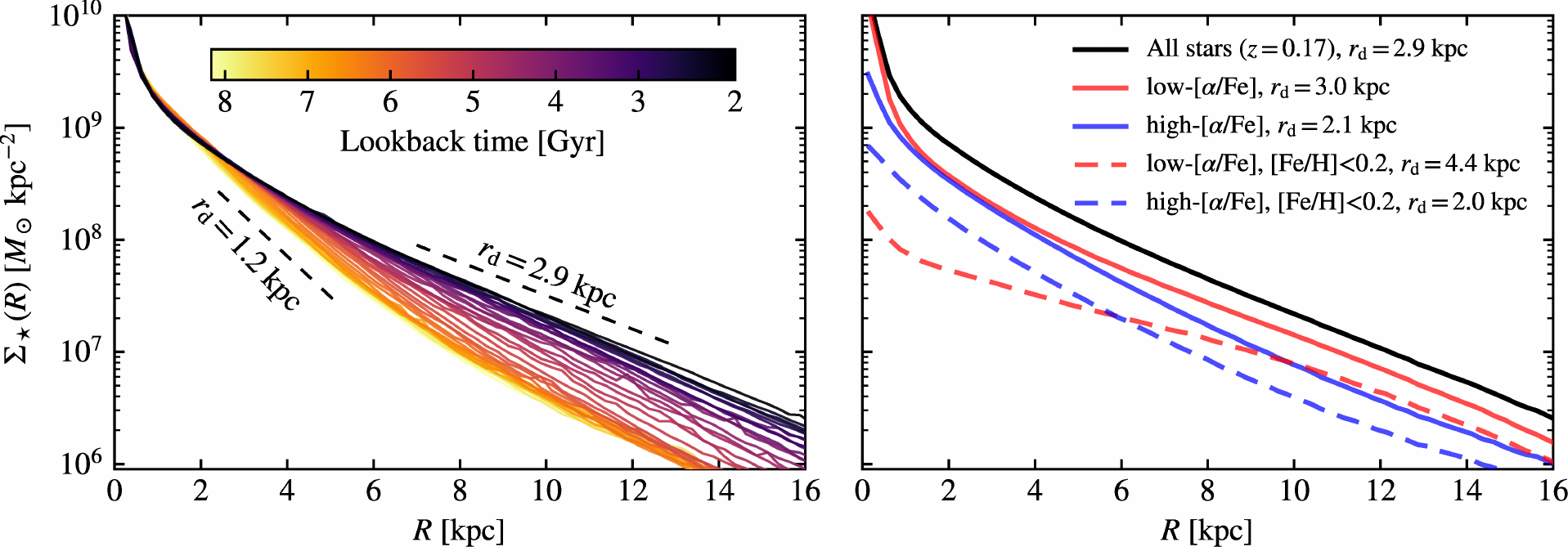}\\
\includegraphics[width=0.83\textwidth]{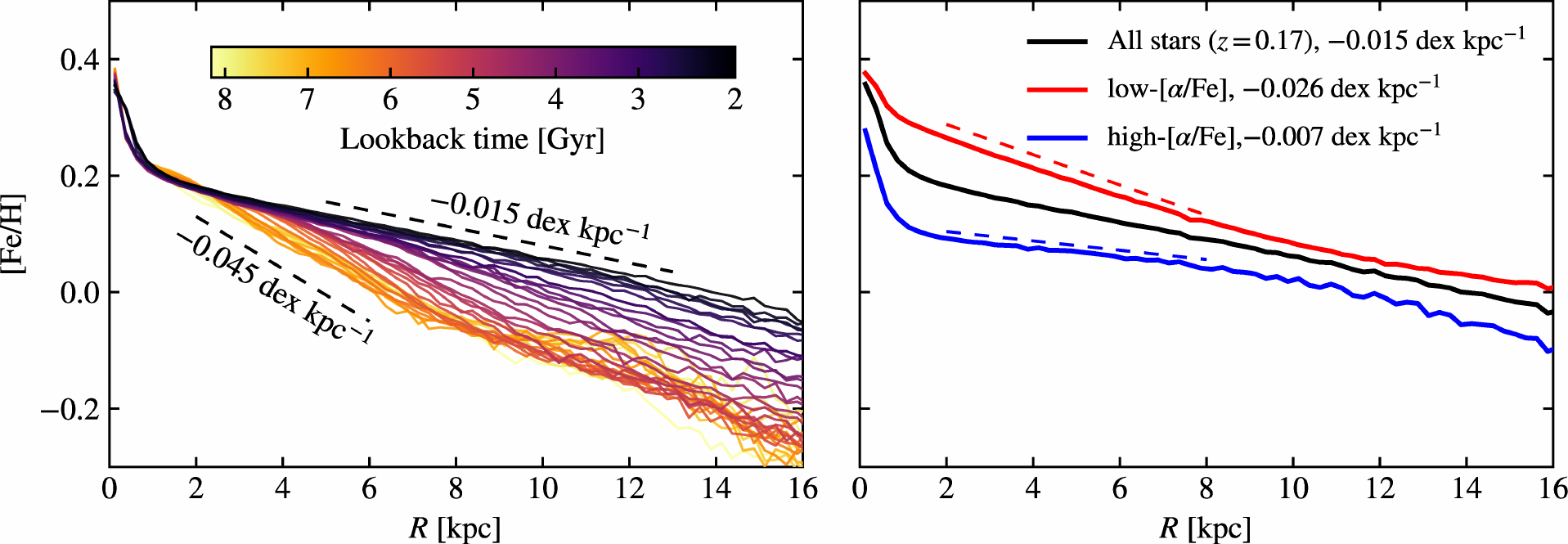}
\caption{Top left: Evolution of stellar surface density profiles from $z=1$ ($t=8\Gyr$). Profiles are computed considering all stars out to $3\kpc$ above the galaxy's mid plane. Top right: $\Sigma_\star(R)$ decomposed into profiles for low-\afe (red solid line) and high-\afe (blue solid line) stars. Also shown are profiles for low-and high-\afe stars (dashed lines) restricted to \feh$<0.2$. Bottom left: Evolution of radial metallicity profiles from $z=1$ ($t=8\Gyr$). Bottom right: \feht-profile at $z=0.17$ (black) decomposed into profiles for low-\afe (red line) and high-\afe stars (blue line). 
}
\label{fig:profileevol}
\end{center}
\end{figure*}

\begin{figure}
\begin{center}
\includegraphics[width=0.43\textwidth]{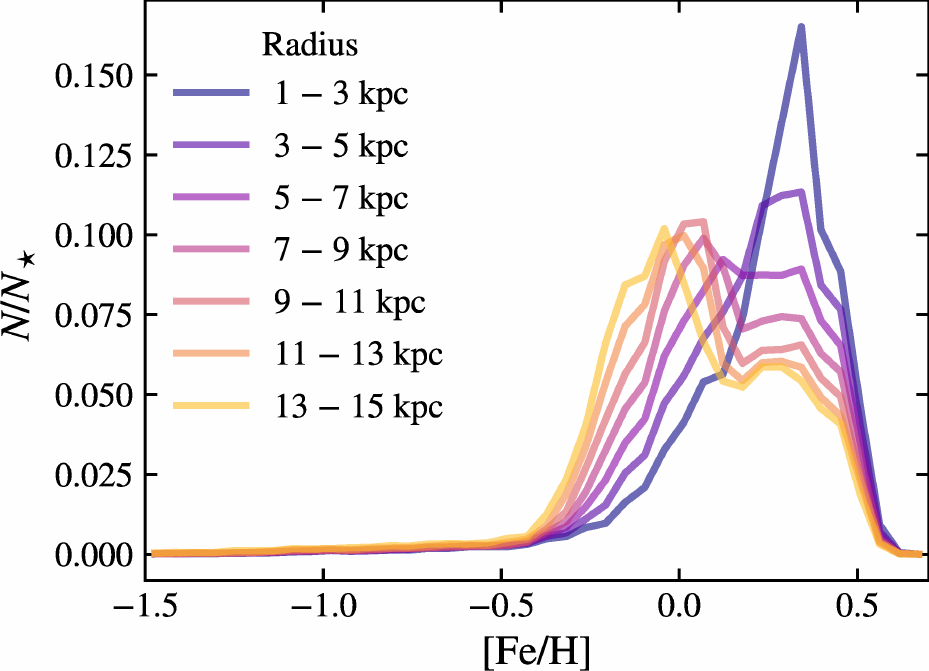}
\caption{Metallicity distribution functions across the galaxy at the current epoch. The negative radial metallicity gradient in Fig.~\ref{fig:profileevol} is reflected in the MDFs, where the inner and outer discs' peak at \feh$\approx 0.3$ and $0$ respectively.
}
\label{fig:MDF}
\end{center}
\end{figure}

We quantify the above chain of events in Fig.~\ref{fig:discprofiles} which shows the evolution of the ISM in terms of its cold ($T<10^4~{\rm K}$) gas surface density ($\Sigma_{\rm gas}$), \afe and \feht. In the local Universe, star formation on kpc-scales is observed to be inefficient for surface densities less than a few $\Msol\pc^{-2}$ \citep[][]{bigiel2008}, which we indicate in red in the top panel. The rapid outer disc formation phase at $z\sim 1-1.5$ is apparent in $\Sigma_{\rm gas}$, with star formation since then becoming possible at $R\gtrsim4\kpc$. At $z=1.3$, the difference in the inner and outer disc's metallicity is striking, with the radial \feht-profile (bottom panel) decreasing by 1 dex from supersolar values at $R\lesssim4\kpc$ to \feh$<-0.5$ in the outer disc, all at the same uniform (low) \afe (middle panel). It is this property that leads to the co-existence of the high-and low-\feh stellar populations discussed in the previous section (bottom panel of Fig.~\ref{fig:chembimodal}). Furthermore, the fact that the ISM at earlier times featured a higher \afe at equally low \feh (dark blue line in Fig.~\ref{fig:discprofiles}) is what allows for a chemical bimodality to exist in the stars. A time series of how the stellar \afe-\feh plane is built up from this process is presented in Appendix~\ref{appendix:evolve}.

We emphasize that even though the emergence of the low-\afe sequence coincides with the epoch of LMM, it is not the infall and mixing of the entire ISM of another galaxy that allows for the bimodality to form \citep[but see e.g.][]{Buck2020b}. In fact, the ISM of the merging galaxy is too metal-rich  (\feh$\gtrsim0$, see Fig.~\ref{fig:largescale} and \citetalias{Renaud2020}) to give rise to the chemical bimodality. Rather, the massive merger is here found to mark a time when significant amounts of gas mass and angular momentum growth takes place due to cosmological cold flows and stripped gas from satellite galaxies. In addition, outflows and galaxy interactions compress the CGM (which is low in \afe and \feht), triggering enhanced gas cooling. This leads to cloud formation via thermal instabilities \citep[][]{Binney2009,Joung2012} and accretion \citep[][]{Hobbs2015} that further promotes the outer disc to grow. The condensation of cold gas from the hot halo is visually apparent in the top right panel of Fig.~\ref{fig:largescale}, and has observational implications in terms of detectability and covering fractions of atomic hydrogen around galaxies \citep[e.g.][]{Rhodin2019}. 

\subsection{Disc growth and secular evolution in the past 8 billion years}
\label{sec:discgrowth}
Having established the mode of high redshift disc assembly, the next sections focus on how the galaxy grows and evolves to the current epoch. This allows us to understand whether or not, and how, the seemingly non-trivial formation scenario relates to the Milky Way and local spiral galaxies.  To bring context to our results we contrast them to observational and theoretical work throughout. 

After the outer disc has formed, the spatial structure of the galaxy is complex. Fig.~\ref{fig:discs2} shows a time sequence of stellar bolometric luminosities, with the galaxy aligned to the disc plane defined by the inner disc ($R<4\kpc$). The outer disc that begins to form $\sim 9\Gyr$ ago defines its own disc plane, with stars forming almost orthogonally to the inner disc. Over subsequent billions of years, the inner and increasingly more massive outer disc gravitationally torque, leading them to align more closely over time. This process eventually allows for coherent spiral structure and radial mixing of stars throughout the entire galaxy. Since a lookback time of $\sim 5\Gyr$, no signs of the initial misalignment can be found, apart from a warp at large radii. In \citetalias{Renaud2020b}, we explore observational consequences of the misaligned discs.

\begin{figure*}
\begin{center}
\includegraphics[width=0.88\textwidth]{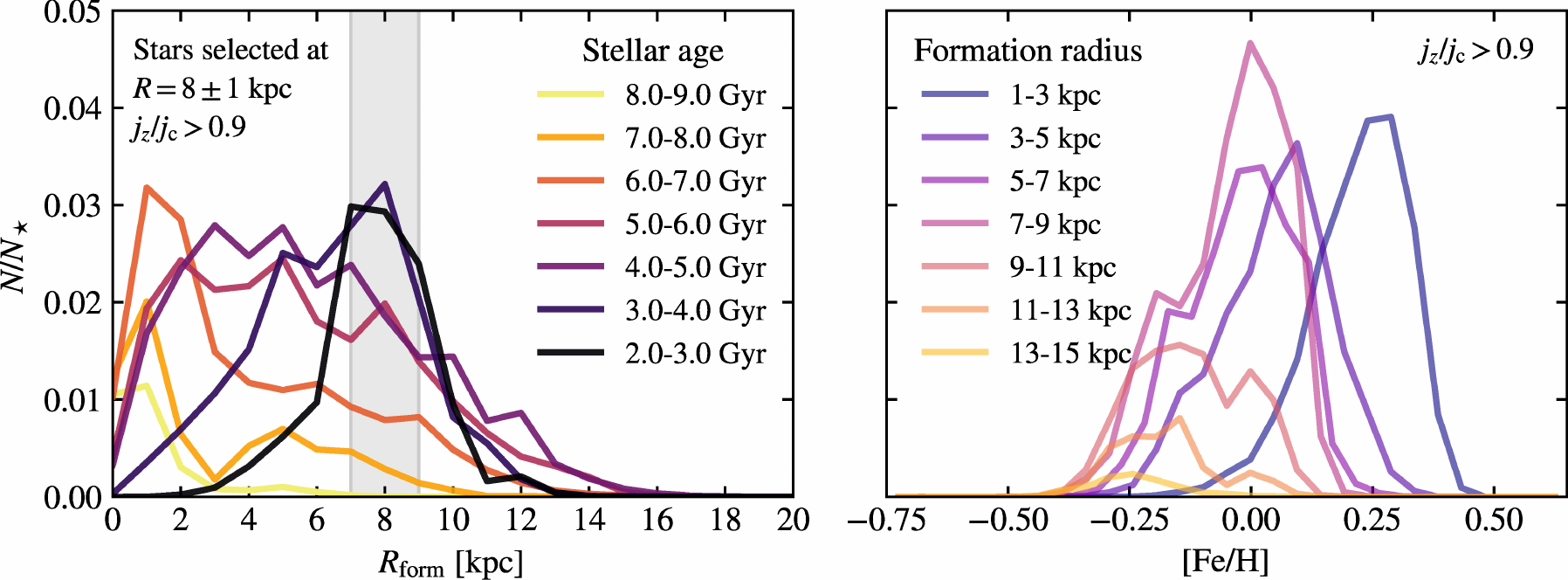}
\caption{Left: Formation radius distributions for stars on nearly circular orbits residing at $R=7-9\kpc$ at the current epoch. The majority of old stars have migrated from the inner disc, in part because the disc beyond 4 kpc was still assembling at $z\sim 1$. Right: MDFs for stars migrating from different galactocentric radii, demonstrating that stars originating from the inner disc have higher metallicities.
}
\label{fig:migration}
\end{center}
\end{figure*}

\subsubsection{Stellar surface density profiles}
\label{sect:profiles}
The top left panel in Fig.~\ref{fig:profileevol} shows the evolution of radial stellar surface density profiles since $z=1$. All profiles are well fitted by single or broken exponential profiles -- a generic feature of stellar scattering \citep[e.g.][]{ElmegreenStruck2013}. The profile inside 2 kpc does not evolve over the entire time span, in agreement with the observationally inferred evolution of current day Milky Way-mass galaxies presented in \citet{vandokkum2013}. In contrast, at larger radii ($>2\kpc$) the galaxy grows smoothly over time, with an exponential scalelength $r_{\rm d}=1.2\kpc$ at $z=1$, increasing monotonically to $r_{\rm d}=2.9\kpc$ at $z=0.17$. In the simulation's Solar vicinity ($R\sim 8.5\kpc$, \citealt{BlandHawthornGerhard2016})
 the surface density is $35\Msol{\rm pc}^{-2}$, close to what is observed in the Milky Way \citep[$\sim 30\Msol{\rm pc}^{-2}$, e.g.][]{Flynn2006,Bovy2012thick}.

In the Milky Way, the observed disc scalelength depends on properties of the underlying stellar populations, in particular the distribution of ages and elemental abundances \citep[e.g.][]{Bovy2016,Mackereth2017}. In the top right panel of Fig.~\ref{fig:profileevol} we explore this concept in \vgs by showing surface densities separately for stars in the low-\afe and high-\afe sequence (split at \afe$=0.39$), i.e. the `chemically defined' thin and thick discs. The young, low-\afe disc is well approximated by an exponential profile with $r_{\rm d}\approx 3.0\kpc$, whereas the older high-\afe disc is smaller with $r_{\rm d}\approx 2.1\kpc$. By restricting the analysis to \feh$<0.2$, hence avoiding stars where the sequences connect in \afe-\feht, the size difference is even greater, with disc scalelengths of $4.4$ and $2.0\kpc$ in the low-and high-\afe sequence, respectively. Such size differences are observed in the Milky Way where the low-\afe thin disc has twice the scalelength of the high-\afe thick disc, with $r_{\rm d}=2$ and $3.8\kpc$ respectively (\citealt{Bensby2011}, see also \citealt{Cheng2012} and \citealt{Bensby2014}). Further sub-division into individual mono abundance populations reveals an entire range of scalelengths in the Milky Way \citep[$2\kpc<r_{\rm d}<4.5\kpc$,][]{Bovy2012mono}, with the largest values recovered for low-\afet, low-\feh stellar populations, akin to the \vgs simulation (dashed red line in Fig.~\ref{fig:profileevol}).

The origin of the thick/thin disc size diversity ultimately relates to the formation time of different components. Tidal torques from large scale structures continuously increase the specific angular momentum content of accreting gas over time \citep[][]{Peebles69,FallEfstathiou80, Pichon2011}. A late formation time hence favours disc formation \citep[e.g.][]{MoMaoWhite98}, but this is not a sufficient criterion; accreting gas must also add constructively to the galaxy's angular momentum reservoir for extended discs to form \citep[][]{Sales2012,Kretschmer2020}. This line of arguments is broadly why the late-time forming low-\afe population is more extended than the high-\afe one, and why the formation of disc galaxies in general proceeds inside-out.

\subsubsection{Radial distribution of metallicities}
\label{sec:metgrad}
The bottom left panel in Fig.~\ref{fig:profileevol} shows the evolution of the radial \feht-profile since $z=1$. Each radial bin is the average \feh of stars out to 3 kpc above the midplane. Akin to the surface density profiles, the metallicity in the inner 2 kpc does not evolve over this time, whereas the metallicity gradient in the outer disc smoothly increases from $\Delta$\feht$/\Delta R=-0.045~{\rm dex}~\kpc^{-1}$ at $z=1$ to $-0.015~{\rm dex}~\kpc^{-1}$ at $z=0.17$. 

In Fig.~\ref{fig:profileevol} we compute \feht-profiles separately for low-\afe and high-\afe stars. A steeper gradient is found for the younger low-\afe population for $R\lesssim 10\kpc$, with $\Delta$\feht$/\Delta R=-0.026~{\rm dex}~\kpc^{-1}$, compared to $-0.007~{\rm dex}~\kpc^{-1}$ for the high-\afe stars. A similar flattening of the profiles occurs when considering older stars, or stars further away from the disc midplane -- a trend observed in the Milky Way \citep[][]{Hayden2014,Anders2017}. This is due to the fact that both in the Milky Way and \vg, high-\afe stars tend to be older and reside in a thick configuration, see Section \ref{sect:vertical} \citep[see also][]{Minchev2014,Ma2017}. 

However, regardless of abundance cuts, the recovered metallicity gradients are smaller than those found in the Milky Way, where $\Delta$\feht$/\Delta R\sim -0.058~{\rm dex}~\kpc^{-1}$ for young stars in the thin disc \citep[][]{Luck2011, Anders2017}. We can explore the reasons for shallower radial metallicity gradients by studying the metallicity distribution function (MDF), shown in Fig.~\ref{fig:MDF}. At all galactocentric radii, the MDFs show a spread in \feh of $\sim$1 dex. The inner disc's MDF peaks at \feh$\approx 0.3$ with few stars at sub-solar metallicities. At gradually larger radii, the contribution from low-\feh stars increases, allowing for a peak metallicity at \feh$\sim -0.1$ to $0$ in the outer disc ($R>7\kpc$). The simulated MDFs are qualitatively similar to those derived for the Milky Way \citep[cf. fig.~5 in][]{Hayden2015}. However, the outmost radial bins in \vgs do not reach as low \feh as is observed in the real Milky Way, a sign of a different enrichment history. 

Furthermore, the simulation features a greater contribution of stars with supersolar metallicity at all radii. To be specific, the peak at \feh$\approx 0.3$~dex found in the inner disc appears as a second, but less prominent, maximum across the entire galaxy, even at $R>13\kpc$. This feature is not observed in the Milky Way's Solar neighbourhood and beyond, and it contributes significantly to the shallower metallicity gradients discussed above. This high \feh cannot, at least in general, be due to local star formation conditions in the outer, more metal-poor, disc. Rather, is arises due to radial migration of metal-rich stars from the inner regions of the galaxy, which we explore next.

\begin{figure*}
\centering
\includegraphics[width=0.98\textwidth]{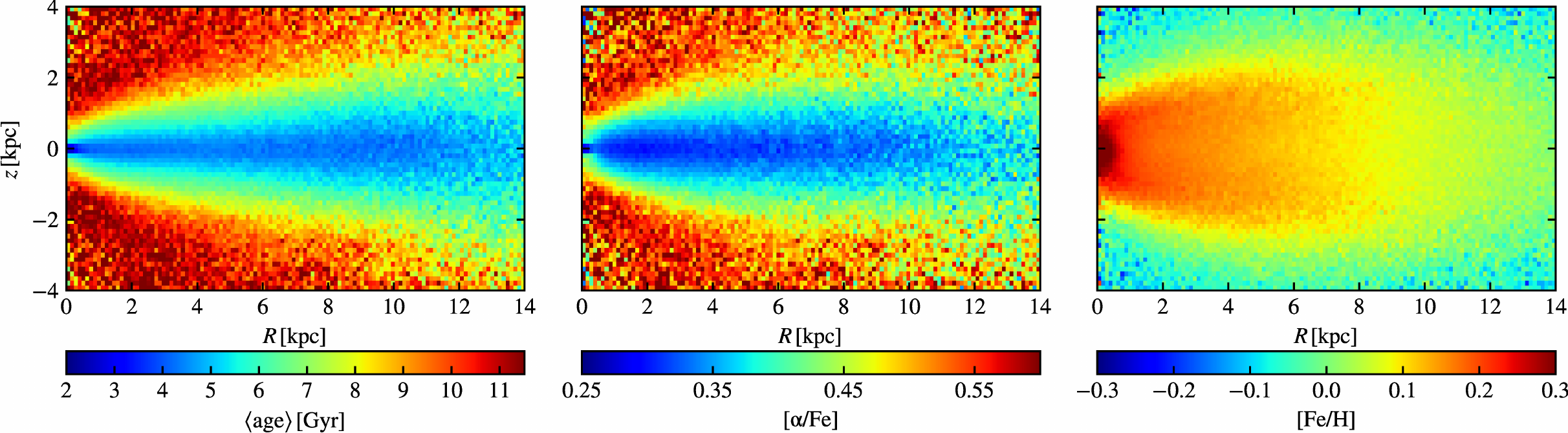} 
\caption{Average stellar ages (left), \afe (middle) and \feh (right) as a function of galactocentric radius and vertical distance from the midplane of \vgs at the final simulation time. Younger stars predominantly reside in a thin flaring disc (left panel), with gradually older (also flaring) stellar populations away from the disc midplane. The \afe structure (middle panel) mirrors the age structure, with a vertical gradient in \afe present at all radii. Low values of \afe are found in the thin younger disc and high values in the thick older disc. The \feh distribution (right panel) shows a negative radial metallicity gradient in the midplane of the disc. At increasing height above the midplane, the profiles are shallower, even becoming positive above $1\kpc$ in the inner disc ($R\lesssim 4\kpc$).
}
\label{fig:discedge}
\end{figure*}

\subsubsection{Radial migration and the Solar neighbourhood}
\label{sec:radmig}
While the galactic mass and metal growth since $z=1$ is completely dominated by in situ star formation (see also \citetalias{Renaud2020}), the radial density and metallicity profiles do not reflect the formation radii of the stars. As discussed in Section \ref{sect:intro}, migration of stars via churning by a bar or (transient) spiral waves \citep[][]{SellwoodBinney2002,Minchev2013,Mikkola2020} can mix stars radially. 

To understand whether radial migration plays a role in \vg, we select at the final simulation time stars around $R=8\pm 1 \kpc$ (the `Solar neighbourhood') out to 2 kpc away from the midplane and track them to their formation positions. We limit our analysis to stars formed at $z<1$, as well as currently being on nearly circular orbits, with $j_z/j_{\rm c}>0.9$, as spiral arm churning preserves orbit circularity. Here $j_z$ is the component of a star particle's specific angular momentum vector parallel to the galaxy's net angular momentum and $j_{\rm c}$ is the specific angular momentum of a circular orbit with the same specific energy as the true orbit, calculated following the approach in \citet{Elbadry2018}. Close to $z=1$, the misaligned nature of the inner and outer discs complicates the notion of a cylindrical formation radius across the galaxy. To mitigate this we instead consider the spherical radius from the center of the galaxy to denote the formation radius.

The left-hand panel of Fig.~\ref{fig:migration} shows formation radius distributions of the selected stars for different ages. A minority of the oldest star ($\gtrsim 6\Gyr$) on nearly circular orbits are born at their current radial location, with the majority having migrated from the inner disc, in agreement with recent theoretical and observational studies of Solar neighbourhood stars \citep[e.g.][]{Minchev2013,Frankel2018,Minchev2018,Feltzing2020}. 
This bias of old stars originating from the inner disc is due to the galaxy beyond 4 kpc still assembling $8\Gyr$ ago \citep[see also][]{Aumer2017}. Subsequent disc growth allows for younger stellar populations to have their formation radius distributions shifted closer to their current radial position. For the youngest stars ($2-3\Gyr$), the distribution is nearly symmetric around $R=8\kpc$, indicating that, at least at late times, inward and outward migration are of equal importance. This is expected for spiral arm churning due to angular momentum conservation \citep[][]{SellwoodBinney2002}.

Migration of older stars from the inner disc entails higher metallicities, as shown in the right-hand panel of Fig.~\ref{fig:migration}; all stars with \feh$>0.25$ at the current Solar radius in \vgs have formed in the inner 5 kpc. This process explains the origin of the high-\feh peaks in the outer disc's MDFs (Fig.~\ref{fig:MDF}). As this feature is not observed in the Milky Way, it indicates that radial migration in \vgs is more efficient than in the real Milky Way, despite the absence of a bar. The migrated high-\feh stars in the outer disc also contribute to the shallower than observed radial metallicity gradients.

The current average metallicity and stellar density around the Solar circle in \vgs is hence a mix of stars of all ages, over 1 dex in \feht, and formation radii across the entire galaxy. However, the fact that the young low-\afe population is more extended than the older high-\afe stars, as also observed in the Milky Way, indicates that radial migration cannot wash out all underlying structure provided by inside-out growth of the galaxy \citep[as also pointed out by][]{Minchev2014}. 

\begin{figure}
\centering
\includegraphics[width=0.42\textwidth]{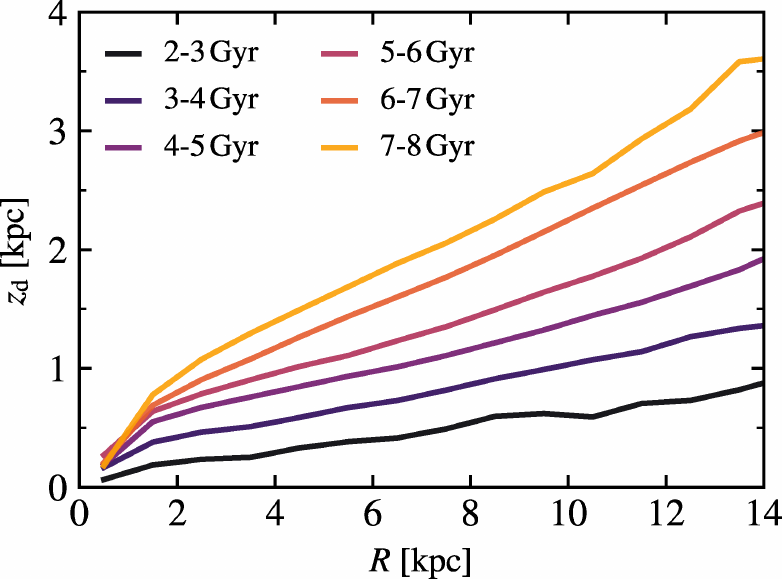} 
\caption{Exponential scale height of stars with different ages as a function of final galactocentric radius. All mono-age populations flare, with increasing scale heights for older stars. 
}
\label{fig:scaleheight}
\end{figure}

\begin{figure*}
\centering
\begin{tabular}{cc}
\includegraphics[width=0.44\textwidth]{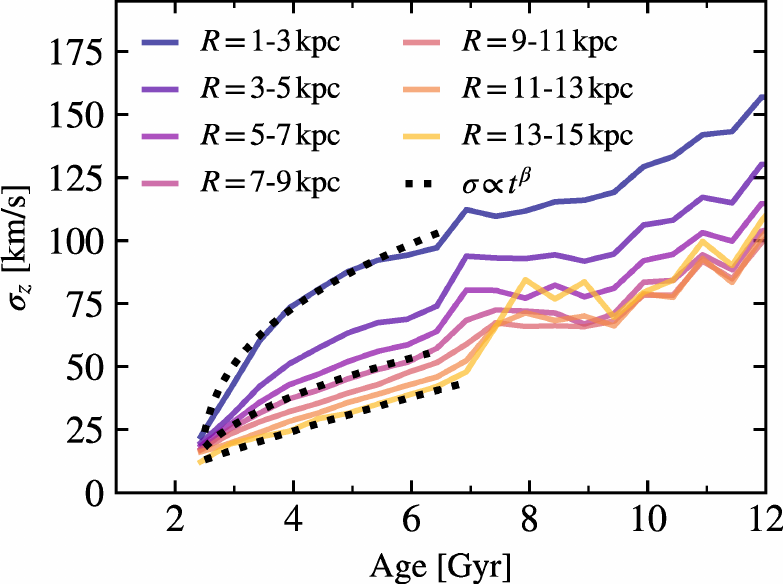} 
\includegraphics[width=0.44\textwidth]{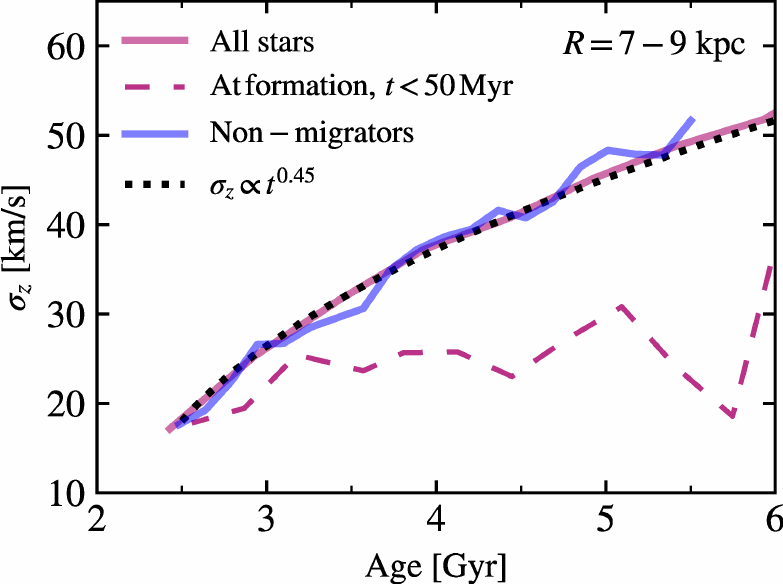} 
\end{tabular}
\caption{Left: The age-(vertical) velocity dispersion relation for stars at different radii. Young stars are kinematically colder at all radii, with a vertical velocity dispersion $\sigma_{\rm z,\star}\sim15\kms$ for newly born stars in the outer disc. Stars in the inner disc always have a higher $\sigma_{z,\star}$ compared to the stars in the outer disc. Dashed lines are fits assuming $\sigma\propto t^{\beta}$. The power law index $\beta$ depends on galactocentric radius, with, from top to bottom, $\beta=0.35, 0.45$ and 0.8. Right: AVR for all stars residing at $R=7-9\kpc$ (gray solid line), power-law fit to this relation (short-dashed line), the velocity dispersion at formation at this location (long-dashed line), and the AVR for stars were born at this location but never migrated (blue solid line).
}
\label{fig:AVR}
\end{figure*}

\subsection{Through thick and thin - vertical structure and kinematics}
\label{sect:vertical}
Having quantified the radial structure and evolution of the disc, we next turn to an analysis of the vertical structure. Fig.~\ref{fig:discedge} shows average stellar ages, \afe and \feh as a function of galactocentric radius and vertical distance from the midplane at the final simulation time. The left panel shows that young stars predominantly reside in a thin flaring disc, with gradually older stellar populations at large distances above the midplane. This is quantified in Fig.~\ref{fig:scaleheight} which shows the exponential scaleheight\footnote{Defined via the vertical density profile $\rho(z)=\rho_0 \exp(-z/z_{\rm d})$} of stars with different ages as a function of galactocentric radius. In the inner kiloparsecs, the large stellar and gas surface densities result in small scaleheights, being $<100$ pc for the youngest age bin. All mono-age populations flare, with monotonically increasing scaleheights for older stars at all radii \citep[see also][]{Minchev2015}. For example, at $R=8\kpc$, stars in the age range 2-3 Gyr have a scaleheight $\approx 500\,{\rm pc}$ which increases to $\approx 2.5\kpc$ for stars with ages of $7-8\Gyr$. The notion that young stellar populations reside in a thin as well as a radially more extended disc (as shown in Section \ref{sect:whathappened}) is commonly referred to as an `inside-out, upside-down' galaxy formation scenario \citep[e.g.][]{Bird2013}.

Returning to Fig.~\ref{fig:discedge}, the \afet-structure in the middle panel mirrors the age structure in the left panel; at all radii a vertical gradient and radial flaring in \afe is present, with low values of \afe found in the thin younger disc and high values in the thick older disc -- the canonical picture of a thin vs. thick disc.

The vertical \feht-distribution in the right hand panel reveals why the radial metallicity gradient (see Section~\ref{sec:metgrad}) depends on \afet, and by extent also to age and height above the disc's midplane; while a negative radial metallicity gradient is present in the midplane of the disc, it flattens with increasing height above the midplane. Above $\sim 1\kpc$ the radial gradient even becomes \emph{positive} in the inner disc ($R<4\kpc$). Again, this a natural outcome of inside-out galaxy growth (the inner galaxy has had more time to enrich) and disc flaring (young low-metallicity stars in the outer disc can be found above the midplane), a property identified in theoretical work by \citet{Miranda2016} and observed in the Milky Way \citep[e.g.][]{Anders2014,Wheeler2020, Wang2020}.

\begin{figure*}
\begin{center}
\includegraphics[width=0.89\textwidth]{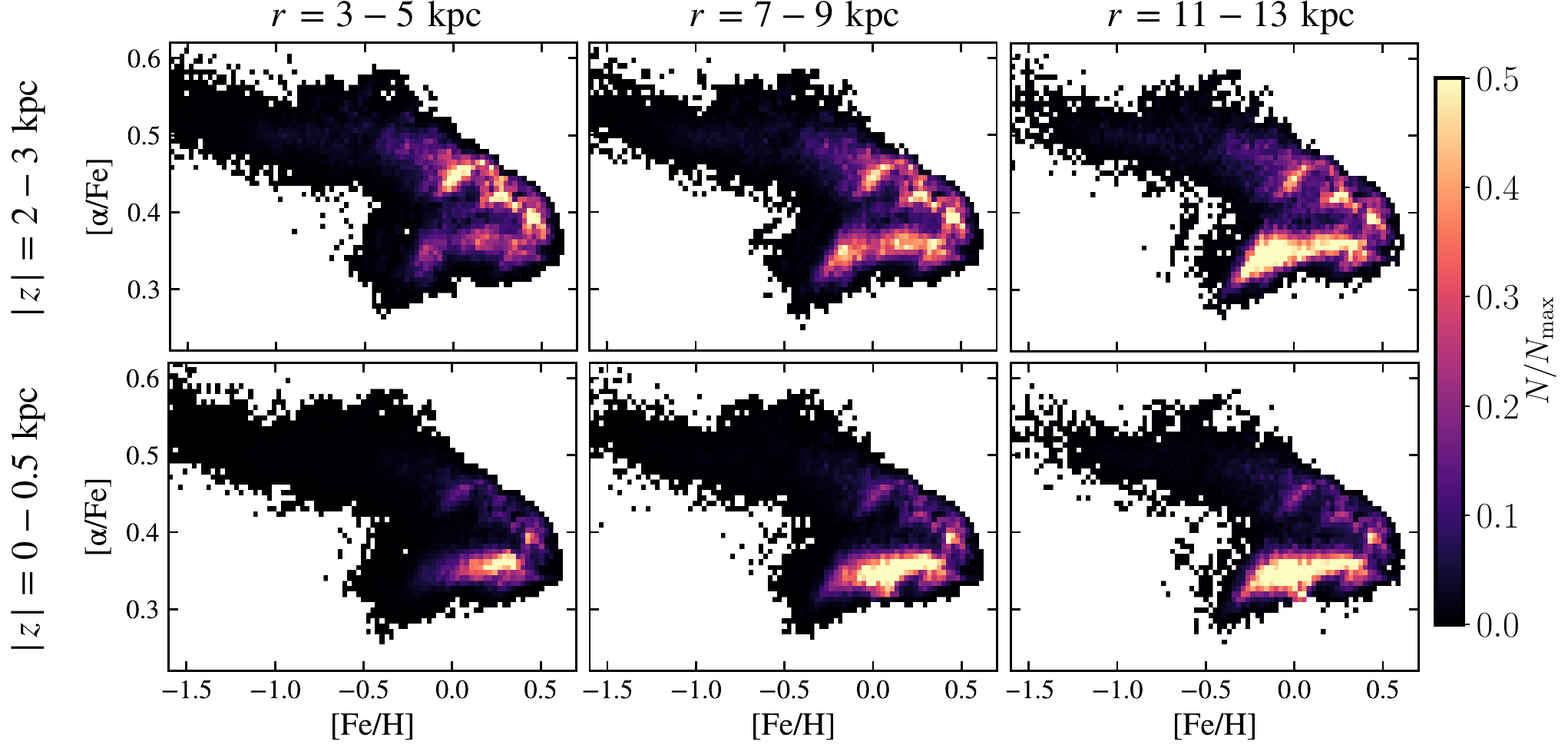} 
\caption{\afet-\feh distribution for a range of radial and vertical locations in the disc. Each panel is normalised using their respective total number of stars. Stars close to the disc midplane ($<0.5\kpc$) are predominantly low-\afet, with  populations in the outer disc featuring a wider range of \feh compared to the inner disc. Above the disc midplane ($2-3\kpc$), the high-\afe sequence contributes significantly at all radii, with the low-\afe sequence gradually contributing more at larger radii.}
\label{fig:alphastruct}
\end{center}
\end{figure*}

\subsubsection{Vertical kinematics and disc heating over cosmic time}
\label{sect:AVR}
The left panel in Fig.~\ref{fig:AVR} shows the age-(vertical) velocity dispersion relation (AVR) for stars at different radii. Across the entire galaxy, young stars are kinematically cold with vertical velocity dispersions $\sigma_{\rm z,\star}\sim15 \kms$ in the outer disc \citep[compatible with Solar neighbourhood stars in the Milky Way,][]{Holmberg2009,Casagrande2011,Yu2018}. The velocity dispersion increases with the age of stellar populations at all radii, meaning the traditional notion of a thick, old, and kinematically hot disc is present. Furthermore, a radial gradient exists at all stellar ages, with stars in the inner disc always having a higher dispersion compared to the stars in the outer disc. We note that this is compatible with disc flaring due the higher surface density in the inner disc\footnote{For a single component isothermal population of stars with an exponential vertical distribution, $\sigma_{z,\star}=(2\pi G\Sigma_{\star} z_{\rm d})^{0.5}$. At $R=12\kpc$, $\Sigma_\star$ is 100 times smaller than at $R=2\kpc$, whereas $z_{\rm d}$ only varies by at most factor of 5 for all stellar ages.}.

A number of notable features exists in the AVR: at stellar ages $>10\Gyr$, stars predominantly have halo-like kinematics with $\sigma_{z,\star}$ always in excess of $75\kms$, an outcome of high redshift merger activity (for more details, see \citetalias{Renaud2020}). For $\sim 7-10\Gyr$, correlated with the LMM, $\sigma_{z,\star}$ is constant at any given radius with thick disc kinematics ($\sigma_{z, \star}\sim 55-75\kms$ for $R>5\kpc$). This feature is caused by the merger itself \citep[see also][]{Buck2020a} as well as star formation in the misaligned, gas rich turbulent disc (for a detailed analysis, see \citetalias{Renaud2020b}).

In the last $\sim 7\Gyr$, mergers become more infrequent, leading to gravitational scattering off molecular clouds playing a more significant role for disc heating \citep[for a review, see][]{Sellwood2014}. As commonly done in the literature, we approximate the AVR by a power-law, $\sigma_{z, \star}\propto t^{\beta}$, and find that the index $\beta$ depends on the radial position in the galaxy. For $R=1-3\kpc$, $\beta\sim 0.35$ with a gradual increase towards larger galactocentric radii ($\beta=0.45$ and 0.8 at $R=7-9\kpc$ and $R=13-15\kpc$, respectively). Such radial differences were highlighted by \citet{Aumer2016} (see their fig.~4) who emphasized the role played by the initial conditions, e.g. mass in thick disc component, of the galaxy. However, $\beta>0.5$ is unphysical in a system where stars are heated by fluctuations that constitute a stationary random process \citep[][]{Wielen1977}, with $\beta = 0.25$ to be expected purely from molecular cloud heating based on analytical models \citep[][]{Lacey1984}. \citet[][also \citealt{Aumer2017}]{Aumer2016} proposed radial migration as an explanation for their models with $\beta>0.5$; at any given radial bin, each coeval stellar population of stars that contributes to the AVR are born over a range of galactocentric radii, and have hence undergone different heating histories. As such, the AVR at any location is not produced by a single stationary heating law. 

To test this hypothesis we compare, in the right-hand panel of Fig.~\ref{fig:AVR}, 1) the AVR for all stars currently residing in the simulation's Solar neighbourhood ($R=7-9\kpc$), 2) the initial (formation) velocity dispersions of all stars at this location, and 3) the AVR for stars that formed at this location but never experienced radial migration. We find that non-migrating stars experience a near identical level of heating as the overall population, indicating that in the \vgs simulation, the AVR does not arise from radial migration. Analogous results were put forward by  \citet{Minchev2012} using non-cosmological simulations of barred spiral galaxies.

It is important to emphasize that the Milky Way has experienced significantly less heating than our simulated galaxy, with an observed AVR well fit by $\beta<0.5$ (\citealt{Holmberg2009}, but see \citealt{Seabroke2007}) and $\sigma_{z,\star}\lesssim 30-40 \kms$ for stars as old as 10 Gyr for $R>6\kpc$ \citep[][]{Mackereth2019}. This is a factor of two lower than in \vgs and hence only compatible with the measured velocity dispersions \emph{at formation}. The origins of this discrepancy is unclear, but can, at least in principle, indicate that the Milky Way had a calmer formation and merger history \citep[for example, velocity dispersions in M31 are significantly higher, with $\sigma_\star\sim90\kms$ at $\sim 4$ Gyr,][]{Dorman2015}. Being in a cosmological context, minor mergers constitute non-stationary events that heat the disc in a different manner than molecular clouds \citep[e.g.][]{TothOstriker1992,Velazquez1999,Kazantzidis2009}. Alternatively, the high velocity dispersions indicate that our simulation, despite high numerical resolution, has experienced non-negligible numerical heating. We return to a discussion of numerical artefacts in Section~\ref{sect:discussion}.

\begin{figure*}
\begin{center}
\includegraphics[width=0.85\textwidth]{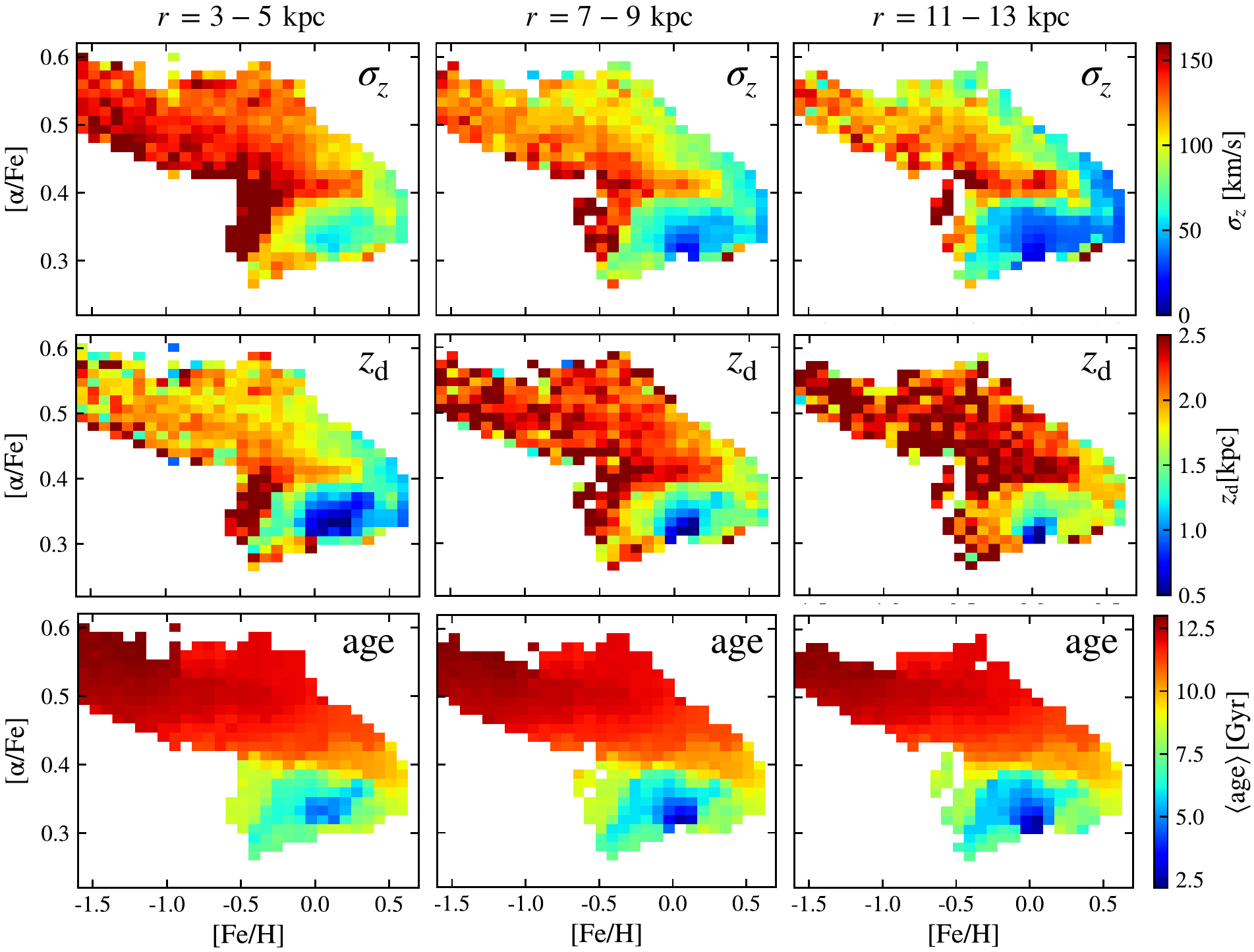} 
\caption{Mono abundance populations colour-coded by vertical velocity dispersion (top row), scaleheight (middle row) and average stellar age (bottom row) at $R=3-5\kpc$ (left), $R=7-9\kpc$ (middle) and $R=11-13\kpc$ (right). Broadly the galaxy features an old, kinematically hot, high-\afe, thick disc, as well as a young, kinematically cold, low-\afe thin disc.
}
\label{fig:monoabundances}
\end{center}
\end{figure*}

\subsection{Properties of mono abundance populations}
\label{sect:MAP}
As discussed in Section~\ref{sect:intro}, spatial, kinematical and chemical properties of individual mono abundance populations can map onto each other in complex manners \citep[][]{RixBovy2013}. To shed light on this, we begin by studying the detectability of the \afet-\feh bimodality as a function of radial and vertical location in the disc, shown in Fig.~\ref{fig:alphastruct}. Each panel is normalised using their respective total number of star particles, allowing for a direct comparison with results from APOGEE in \citet[][]{Hayden2015}. Regardless of location in the disc, a small number of stars always populate the entire disc's range in \feh and \afe -- an outcome of radial migration and mergers. As shown in Section~\ref{sect:vertical}, stars in the structural thin disc (within $<0.5\kpc$) are predominantly found to have low \afet. The negative radial \feh gradient manifests itself by the inner disc ($R<5\kpc$) having stars narrowly distributed around \feh$\sim 0.25$, whereas populations in the outer disc ($R>11\kpc$) extend to \feh$\sim -0.5$. 

Above the disc's midplane ($2-3\kpc$), the high-\afe sequence is present at all radii, with the low-\afe sequence contributing more and more towards the outer disc. At $R>11\kpc$ the low-\afe sequence, i.e. the `chemical thin disc', even dominates in terms of number of stars, a signature also observed in the Milky Way, see for example fig.~4 in \citet{Hayden2015}. In \vg, these trends are a natural consequence of inside-out, up-side down growth (i.e. the age-velocity dispersion and age-scaleheight relations) coupled with disc flaring \citep[see also][]{Bird2013,Minchev2014}, which allows for young low-\afe stars to exist several kpc above the disc's midplane. This property is also observed in the outer disc of the Milky Way \citep[e.g. by LAMOST,][]{Wang2020,Huang2020}.

\subsubsection{How distinct are disc components?}
Fig.~\ref{fig:monoabundances} illustrates how the notion of thin and thick discs becomes nuanced when considering velocity dispersions (top row), scaleheights (middle row) and average stellar ages (bottom row) for mono abundance populations at different galactocentric radii. In broad strokes, the galaxy features an old, kinematically hot, high-\afet, thick disc, as well as a young, kinematically cold, low-\afe thin disc. However, the transition between these components is not always discontinuous in the \afet-\feh plane. For example, at $R=7-9\kpc$, low-\feht, low-\afe populations, which are the first stars to form in the outer detached metal-poor disc at $z\sim 1.5$, and populations at the same \feht, but with high-\afet, all have scaleheights $\gtrsim 2\kpc$ -- a smooth transition between the traditional (chemical) thin and thick disc divide and in qualitative agreement with the Milky Way \citep[e.g.][]{Bovy2012mono}.

A notable feature can be found in the high-\feh part of the high-\afe distribution in the same figure. These stars formed 8-10 Gyr ago, around the time of the LMM, and kinematically they are as cold as the thin disc, with $\sigma_{\rm z,\star}< 40\kms$ at $R=11-13\kpc$. They are hence well separated kinematically from stars with halo-like kinematics ($\sigma_{\rm z,\star}> 100\kms$) at lower \feh in the high-\afe sequence. These stars formed in situ at the time when the galaxy transitioned from its high redshift mode of star formation (frequent mergers, high levels of turbulence, massive star forming clumps, see \citetalias{Renaud2020}), to a more quiescent mode of star formation persisting to the current epoch.

Finally, average stellar ages are well separated in terms of \afe (see bottom row in Fig.~\ref{fig:monoabundances}), confirming the salient dichotomy of an old $\alpha$-enhanced thick disc and a young $\alpha$-poor thin disc. However, the situation for \feh is more complex. Indeed, in the low-\afe population, at all considered radii, \vgs features a wide range of \feh for stars born $\sim 8-9\Gyr$ ago. This is a direct outcome of the galaxy formation scenario described in Section~\ref{sect:whathappened}, and interestingly a property also observed in the Milky Way, albeit not at the precise stellar ages recovered in the simulation \citep[see fig.~7 in][]{Feuillet2019}.

\section{Discussion}
\label{sect:discussion}
We propose a formation scenario for chemically, kinematically and structurally diverse disc components that, despite a seemingly tumultuous origin at $z>1$, leads to a galaxy with current day properties with much in common with those observed in the Milky Way. We next turn to a discussion on how these results compare to other formation channels suggested in the literature, how numerical issues may influence our findings, and whether the proposed chain of events can be supported by observations.

\subsection{Comparison to other models} \label{sect:complit}
The formation scenario at $z>1$ shares similarities to previous work
in the literature. For example, the concept of high redshift gas accretion is central to the classic `two-infall' scenario \citep[e.g.][]{Chiappini1997}. However, in contrast to such analytical models, it is not the dilution of the preexisting ISM by infalling gas that takes place, but rather the settling of gas into a chemically distinct outer disc. The original two-infall model has problems explaining the high-\feh part of the high-\afe sequence, although recent revisions have improved on the match by postulating a later infall of gas, closer to a lookback time of $\sim 9\Gyr$ \citep{Spitoni2019}, the same epoch as identified in \vgs. 

Recent work using cosmological simulations have proposed a number of formation scenarios for the \afet-\feh dichotomy. Using the {\small AURIGA} cosmological simulations, \citet{Grand2018} found distinct \afe sequences in 6 out of 30 galaxies forming in Milky Way mass haloes and identified two pathways: 1) an early ($z>1$) and intense high-\afe star formation phase in the inner region ($R<5\kpc$) induced by gas-rich mergers, followed by more quiescent low-\afe star formation, and 2) an early phase of high-\afe star formation in an outer disc followed by a shrinking of the gas disc owing to a temporarily lowered gas accretion rate, after which disc growth resumes.  Aspects of these scenarios agree broadly with our work, e.g. an early formation epoch of the high-\afe sequence, a late time low-\afe sequence in an extended disc. However, in contrast to our results and the Milky Way, their simulations do not feature any significant overlap between the low-and high-\afe sequences in terms of \feht. Furthermore, their simulated low-\afe sequences do not extend to \feht$<0$ (see e.g. their fig.~1) at any galactocentric radius, in contrast to what is observed in the Milky Way.

\cite{Buck2020b} found bimodal \afet-\feh sequences in 4 out of 6 Milky Way-mass galaxies in the {\small NIHAO\_UHD} simulations suite \citep[][]{Buck2020a}. They argued that it is the dilution of the ISM by a gas-rich merger that allows the disc to lower its \feh while transitioning from high to low-\afe \citep[see also][]{Brook2012thinthick}. As discussed above, in \vgs it is not the major merger itself that adds the metal-poor gas, nor does the bimodality form due to dilution of the main progenitor's ISM. Rather, it is cosmological filamentary accretion that gives rise to an outer metal-poor, low-\afe disc. In agreement with our work, \cite{Buck2020b} found radial migration to be an important mechanism for shaping the galaxy's spatial and chemical structure since $z\sim 1$. As such, radial migration only influences the actual detectability of distinct \afet-\feh sequences at specific radial location across the disc.

\citet{Clarke2019} used a non-cosmological simulation of an isolated galaxy to demonstrate that a chemically bimodal galaxy can arise due to the clumpy nature of star formation, typical in gas-rich high redshift galaxies. In this picture, star formation in massive clumps is rapid and occurs mainly in recently released core collapse SNe ejecta which gives rise to the high-\afe sequence, with more distributed star formation producing the low-\afe sequence. This process is not the main reason for a bimodality in \afe in \vg, but it has nonetheless a distinct imprint in the \afet-\feh plane which we explore further in \citetalias{Renaud2020}.

\begin{figure}
\centering
\includegraphics[width=0.45\textwidth]{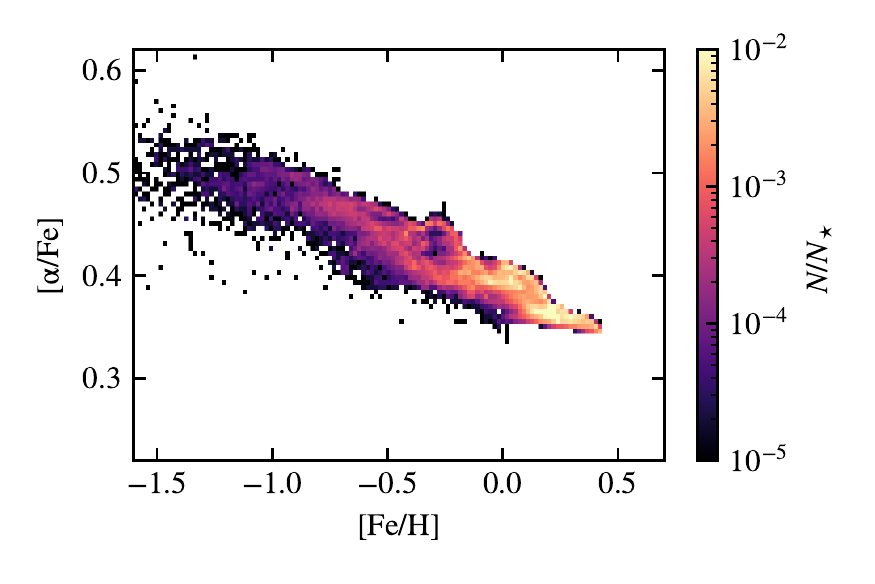} 
\caption{Disc-wide \afet-\feh for \vgs resimulated at a low numerical resolution (finest cell size $\sim 500\pc$) analogous to large volume simulations.}
\label{fig:lowres}
\end{figure}

In contrast to the above studies, \citet{Mackereth2018} found signatures of distinct \afe sequences in only a few per cent out of 133 Milky Way-mass disc galaxies in the {\small EAGLE} simulation volume \citep[][]{Schaye2015}. While it is unknown if such scarcity is to be expected in the real Universe, the mixed results and varying detection frequencies in the literature highlights a central issue; without a clear understanding of the formation channels for chemically distinct discs, we do not yet know the physics and numerical resolution required to capture them.

\subsection{Numerical issues and model uncertainties}
\label{sect:numuncert}
To test the sensitivity of our results to numerical resolution, we resimulated \vgs with resolution (dark matter particle masses $\sim10^6\Msol$, spatial resolution $\sim 0.5\kpc$) and subgrid galaxy formation physics similar to the large volume cosmological simulation {\small EAGLE}\footnote{We adopted the same pressure floor parametrization and restricted stellar feedback to only SNe. The energy released per SN explosion was set to depend on gas metallicity in an identical way to {\small EAGLE}.} \citep[][]{Schaye2015}. We emphasize that this only serves as an illustration, as the subgrid physics in large volume simulations have been designed and carefully tuned to reproduce specific observables (e.g. galaxy luminosity functions) for a given hydrodynamic scheme and adopted numerical resolution.

While global properties of the $z=0$ galaxy, e.g. stellar disc mass and size, match the high resolution simulation, the disc-wide \afet-\feh distribution now only features a single sequence, as shown in Fig.~\ref{fig:lowres}. The gas density distribution and porosity of the ISM and CGM are highly resolution-sensitive, meaning that the way in which gas accretes onto the galaxy, how feedback ejecta propagates, and how gas mixes change with resolution \citep[e.g.][]{Ohlin2019}. In the low resolution version of \vg, cold gas accreting along filaments at $z\sim 1.5$ is  found to rapidly dissolve in the poorly resolved CGM (in contrast to the high resolution case). This in turn alters the way in which gas reaches the galaxy, and prevents a chemically distinct outer disc from forming. The fraction of chemically bimodal galaxies in large volume simulations \citep[e.g.][]{Mackereth2018} would likely increase at higher numerical resolution.

Contributing to the diversity of results from galaxy formation simulations is the impact of the adopted subgrid physics and stellar yields. It is today recognised that efficient stellar feedback prescriptions must be included in cosmological simulations \citep[for a review, see][]{NaabOstriker2017}. This allows for simulated galaxy stellar masses to be compatible with the low galaxy formation efficiency predicted by e.g. abundance matching \citep[e.g.][]{Moster2010}. However, \citet{Gibson2013} demonstrated how some strong feedback models \citep[specifically the prescription adopted in the MaGICC suite,][]{Stinson2013}, can erase all signatures of a chemical bimodality. The manner in which alternative approaches to stellar feedback, e.g. cosmic rays \citep[][]{Booth2013} and runaway stars \citep[][]{Andersson2020}, that are less disruptive to the cold ISM, affect the chemical structure of galaxies is an  interesting topic for future work.

Furthermore, the \afet-\feh distribution is sensitive to core collapse SNe and SNIa rates \citep[][]{Marinacci2014b,Mackereth2018}. These in turn depend on the underlying IMF \citep[][]{Gutcke2019} and stellar binary fraction that cannot be captured in cosmological simulations. Uncertainties associated with parameter choices will propagate into chemical properties of stellar populations \citep[for a detailed discussion, see][]{Philcox2018}. \vgs itself is a good example of this, as it features stars with \afe systematically higher than the Milky Way's by a factor of $\sim 2$. This issue has been identified in other recent simulation efforts \citep[e.g. IllustrisTNG,][]{Naiman2018} and shown to, at least in part, be mitigated by higher, empirically motivated SNIa rates \citep[][]{Maoz2014}. Tests of \vgs with higher SNIa rates, run at low resolution, confirm this picture, which we will return to in future work.

The structure and kinematics of galactic discs are also impacted by numerical resolution, a topic that has been studied for many decades. For example, two-body relaxation in galaxy simulations are known to artificially heat stars \citep[e.g.][]{Sellwood2013}, with $N$-body experiments indicating that over $10^6$ particles are required in the disc alone to mitigate such effects \citep[e.g.][]{Solway2012}. Coupled to this is the role of force resolution in suppressing gravitational instabilities \citep[for analytical work, see][]{romeo94,Romeo1997}, which in turn changes spiral wave propagation, the strength of radial migration and stellar scattering. \vgs features $\sim 10^7$ star particles in the disc, but even so we cannot rule out numerical heating and note that the simulation's stellar velocity dispersions are larger than the Milky Way's by at least a factor of two (see Section \ref{sect:AVR}). In fact, as the simulation's merger history is comparable to what we know about the Milky Way's history \citep[i.e. no major merger in the past $8-10\Gyr$,][]{Ruchti2015}, a numerical origin is likely. The only other galaxy for which an observational estimate of the the stellar velocity dispersion as a function of age exists is M31. This galaxy features a kinematically hotter stellar disc and higher inferred heating rate, with a velocity dispersion of $\sim90\kms$ for $\sim 4\Gyr$ old stars \citep[][]{Dorman2015}, likely due to a recent significant merger \citep[$\sim 2\Gyr$ ago,][]{DSouza2018}.

Currently, cosmological simulations tend to not reproduce the low vertical velocity dispersion observed in the Milky Way's Solar neighbourhood (e.g. \citealt{House2011}, \citealt{Sanderson2020}, but see \citealt{Bird2020}), where $\sigma_{z,\star}\lesssim 20\kms$ over the past $8\Gyr$ \citep[][]{Holmberg2009}. It is safe to claim that the required force and mass resolution for cosmological simulations to robustly capture the interaction between stars, the cold ISM, and stellar feedback processes is not understood, and could be more demanding than what is known from pure $N$-body simulations.  

\subsection{Observational support for the outer disc formation scenario}
The origins of a chemically bimodal galaxy from a rapidly forming, misaligned outer gas disc (surrounding the older inner disc) $8-9\Gyr$ ago is a central prediction from our simulation. We predict that the outer disc only forms low-\afe stars, with the high-\afe sequence originating either in the inner disc or from accreted satellites (see \citetalias{Renaud2020}). Moreover, we predict that $8-9\Gyr$ ago, low-\feht, low-\afe stars in the outer disc formed simultaneously with high-\feh stars in the inner disc. We next discuss these trends in the context of the Milky Way, and whether there are any observational imprints in support of this scenario.

\citet{Bensby2014} conducted a high-resolution spectroscopic study of 714 F and G dwarf and subgiant stars in the Solar neighbourhood. They identified distinct sequences in \afet-\feht, separated around $8-9\Gyr$ ago (see their fig.~21). An inner-outer disc dichotomy was also identified, with the $\alpha$-enhanced population found to have orbital parameters compatible with being born in the inner Galactic disc, and the low-\afe stars mainly coming from the outer disc, akin to what is found in the \vgs simulation.

\cite{Haywood2013} outlined observational support of a two-phase formation history including the existence of a \emph{separate} outer disc at $z\gtrsim 1$. Based on an analysis of 1111 FGK stars in the Solar neighbourhood, they argued that metal-poor thin disc stars in the Solar vicinity have properties best explained by them originating in an outer disc. These stars can be as old as the youngest thick disc stars ($9-10\Gyr$ in their study), indicating that such an outer (thin) disc may have started to form while the thick disc was still forming stars in the inner parts of the Galaxy. This coeval formation scenario is in line with what we find in \vgs.
 
Follow-up work by \citet{Haywood2019} refined this scenario and presented arguments for a partitioning of the Milky Way disc $7-9\Gyr$ ago into an inner and outer region characterized by different chemical evolution. Details of their model differ from ours e.g. they propose the outer Lindblad resonance as a divider of inner and outer regions whereas  such a separation arises due to misaligned gas accretion in \vg. However, it is encouraging that their schematic predictions for the \afet-\feht, age-metallicity and age-\afe relations are closely aligned with ours (see their fig.~6), specifically the rapid development of a low-\feh stellar population in an extended disc $\sim 9\Gyr$ ago.

More support for the rapid formation of an outer disc was presented by \citet{Ciuca2020} who derived stellar ages from {\small APOGEE} using a machine learning approach. They highlighted, in agreement with \vg, the simultaneous formation of low-and high-\feh stars in the low-\afe sequence approximately at the same time as the galaxy transitions from the high to low-\afe sequence. The same feature was also found in earlier work by \cite{Feuillet2019}, with the signature being especially prominent in the outer disc of the Milky Way (see their fig.~3). It is plausible that this signature originates from the very same mechanism identified in this work, which we explore in \citetalias{Renaud2020b}. 

Finally, we re-emphasize that the formation scenario is a non-trivial prediction of our simulation; while broadly arising from gas infall in the early Universe, details of ages and metallicities for mono abundance populations across the galaxy depend on the merger history, angular momentum and mixing of infalling gas, previous enrichment history, interaction-triggered star formation etc. In addition, vertical heating and radial migration in the past 8 Gyr `filters' the formation signal, allowing a mix of stellar populations with different formation histories to be observed in the Solar vicinity. As such, improved stellar ages, for a larger population of stars, will be important for the ability of future joint observational and theoretical work to constrain the formation history of the Milky Way galaxy. Surveys like those carried out with {\small 4MOST} will be instrumental in providing relevant data \citep[e.g.][]{Bensby2019, Chiappini2019}.

\section{Conclusions}
\label{sect:conclusion}
In this work, we have used a new high-resolution cosmological zoom simulation of a Milky Way-mass galaxy, \vg, to understand the origins of chemical, kinematical and structural thick and thin stellar discs. We have demonstrated that \vgs conforms to a number of observed characteristics of the Milky Way (and disc galaxies of similar mass), including its size, gas fraction, stellar surface density profile, rotation curve and star formation history. This agreement motivates our detailed study of the formation and evolution of its internal structure, with a particular focus on the origins of distinct sequences in \afet-\feht, a dichotomy that is well established in the Milky Way. 
Our main conclusions can be summarized as follows:

\begin{itemize}
\item At a lookback time $\gtrsim 9$ Gyr, mergers are frequent and the galaxy is compact, gas rich and turbulent. All stars formed up to this point belong to the high-\afe sequence due to efficiently mixed and recycled core collapse supernova ejecta in the main progenitor and its merging constituents. Supernova type Ia enrichment leads to a gradual lowering of \afe at increasingly high metallicities, with \feh$\sim 0.5$ reached in the innermost part of the galaxy. The contributions from in situ and accreted material, and the roles played by mergers, are presented in \citetalias{Renaud2020}.

\item In connection with the last major merger ($\sim8-9\Gyr$ ago), cosmological inflow of low metallicity gas along filaments and gas from stripped dwarf galaxies lead to a rapid ($<0.5\Gyr$) buildup of an extended ($\sim 10\kpc$) metal-poor outer gas disc around the inner compact ($\lesssim 4\kpc$) metal-rich galaxy. This event leads to low-\afe stars forming simultaneously over a wide range of metallicities ($-0.7\lesssim$\feh$\lesssim 0.5$), as observed in the Milky Way. These values overlap with those of the older high-\afe sequence, leading to the formation of a chemically bimodal galaxy. In \citetalias{Renaud2020b}, we perform an in-depth analysis of how the outer gas disc forms, and the physical processes that trigger it to form stars.

\item The outer star forming disc, formed $8-9\Gyr$ ago, is initially misaligned with the inner one, allowing it to evolve independently over the $\sim 3\Gyr$ of years it takes for gravitational torques to align them. Despite the non-trivial origins, the galaxy transitions into a Milky Way-like disc galaxy featuring, in broad strokes, an old, kinematically hot, high-\afet, thick disc, as well as a young, kinematically cold, low-\afe thin disc. 

\item The galaxy grows inside-out and evolves secularly in the last $\sim8\Gyr$, with its stellar surface density profile well fitted by single or broken exponentials at all times. Close to $z=0$ the exponential scalelength matches the Milky Way's ($\approx 2.9\kpc$), with a size dichotomy for the `chemically defined' thin and thick discs; the low-\afe disc forms late ($z<1$) from high angular momentum material and has twice the scalelength ($\approx 4~\kpc$) compared to the older high-\afe disc ($2\kpc$), in line with the Milky Way.

\item Radial migration shapes the Solar neighbourhood (galactocentric radius $\sim 8\kpc$) metallicity distribution function. Coupled with inside-out galaxy growth, radial migration preferentially redistributes metal-rich (\feh$>0$) stars from the inner galaxy to the outer. \vgs  features a higher than observed fraction of high-\feh stars in the Solar vicinity, possibly indicating a lesser role of radial migration in the Milky Way.

\item The final galaxy's disc scaleheight increases monotonically with the age of the stellar population. This `upside-down' formation scenario (young stars residing in a thin disc), together with the fact that all mono-age stellar populations flare, explains a number of structural features also observed in the Milky Way. Two specific examples are the existence of 1) shallower, or even inverted, \feht-profiles above the midplane, and 2) young low-\afe stars residing in a structurally thick disc beyond the Solar radius.

\item Stars are born kinematically cold, with vertical velocity dispersions $\sim 15~\kms$. Velocity dispersions increase with age and decreasing galactocentric radius due to secular heating process such as gravitational scattering off clouds in the disc.
 Radial migration is not found to affect the age-velocity dispersion relation in the Solar neighbourhood, with stars not experiencing any migration being heated just as much as the general population. However, simulations with higher resolution are necessary to confirm this result. 

\item While a thin--thick disc dichotomy broadly is in place, velocity dispersions, scaleheights and average stellar ages of mono-abundance populations can relate to each other in complex manners depending on location in the \afet-\feh plane. For example, the most metal-poor stars (\feh$<-0.5$) in the low-\afe sequence and old stars in the high-\afe sequence both feature scaleheights $\gtrsim 2\kpc$.

\end{itemize}

The proposed formation scenario of chemically, kinematically and structurally distinct disc components hence leads to a galaxy with current day properties with much in common with the Milky Way. Central to this scenario is 1) the formation of an outer disc from cosmological accretion around the epoch of the last major merger \citep[see also][]{Kretschmer2020} which allows for a chemical bimodality, and 2) secular evolution that shapes the internal structure and leads to many of the commonly observed thin vs. thick disc characteristics. How common such a chain of events are for Milky Way-mass disc galaxies will be explored in future work. 

The forensic evidence necessary to unravel the formation scenario of the Milky Way will only improve in the near future, with a wealth of new data becoming available from upcoming large ground-based spectroscopic surveys, such as WEAVE \citep[][]{Weave2012} and 4MOST \citep[][]{4MOST2019}, together with forthcoming data releases from the astrometric satellite Gaia \citep[]{Gaia2016}.

\section*{acknowledgments}
OA thanks Romain Teyssier, Andrea Macci\`o and Tobias Buck for discussions.
OA, FR, EA and MR acknowledge support from the Knut and Alice Wallenberg Foundation and the Royal Physiographic Society of Lund. OA, FR and EA acknowledge support from the Royal Physiographic Society of Lund. OA is supported by the grant 2014-5791 from the Swedish Research Council. TB is supported by the grant 2018-04857 from the Swedish Research Council. SF and DF are supported by the grant 2016-03412 from the Swedish Research Council. This work used the COSMA Data Centric system at Durham University, operated by the Institute for Computational Cosmology on behalf of the STFC DiRAC HPC Facility (www.dirac.ac.uk). This equipment was funded by a BIS National E-infrastructure capital grant ST/K00042X/1, DiRAC Operations grant ST/K003267/1 and Durham University. DiRAC is part of the National E-Infrastructure. 

\section*{Data availability}
The data underlying this article will be shared on reasonable request to the corresponding author.
 
\footnotesize{
\bibliographystyle{mnras}
\bibliography{vg1.bbl}

}

\appendix

\section{Evolution of the \afet-\feh plane}
\label{appendix:evolve}
\begin{figure*}
\centering
\includegraphics[width=0.92\textwidth]{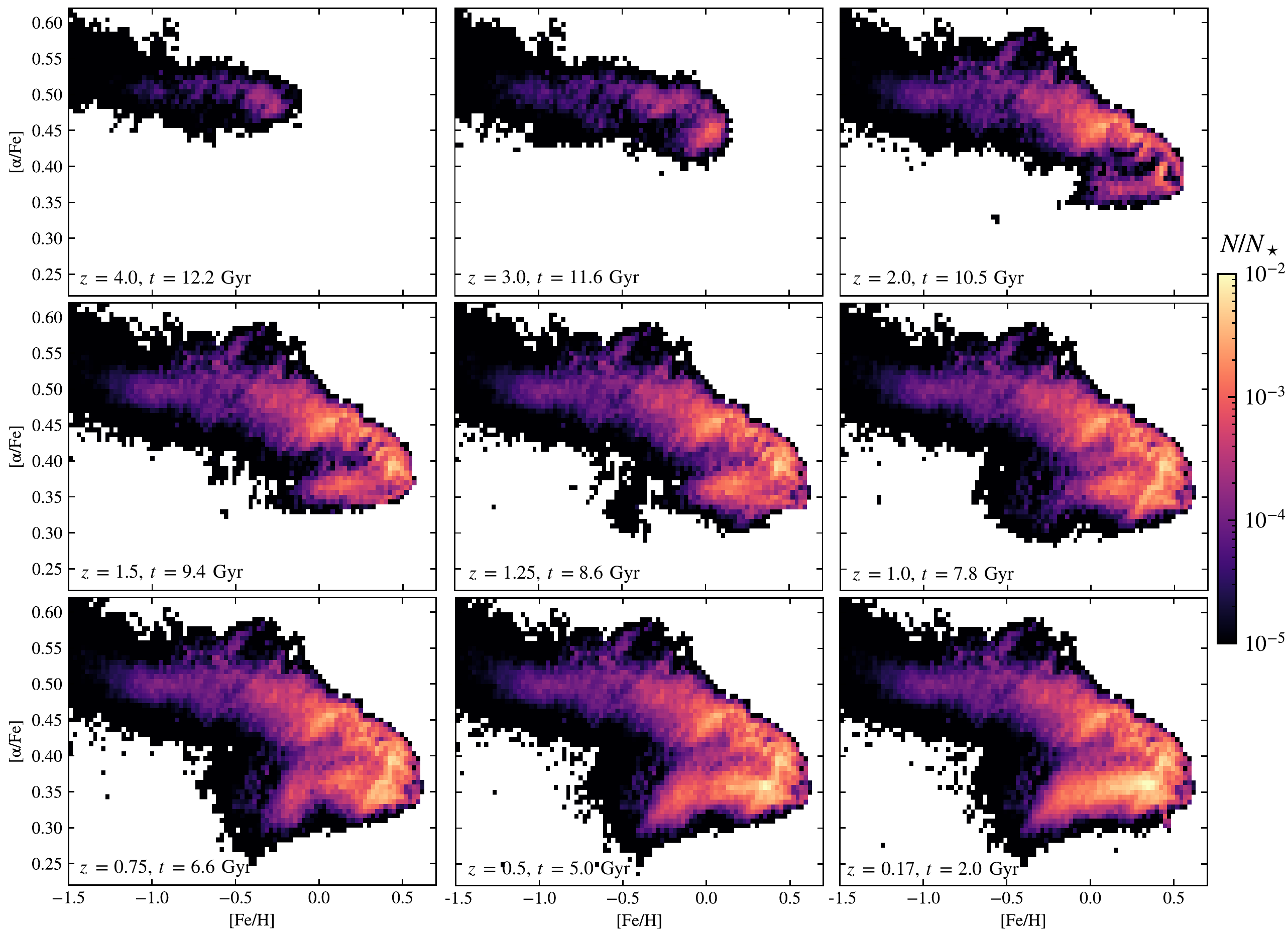} 
\caption{Evolution of \afet-\feh for stars residing at galactocentric radii $<20\kpc$ and $<3\kpc$ from the midplane. The simulation time is indicated in each panel. The 2D histograms show the number of stars ($N$) in each pixel normalized to the number of stars in the entire galaxy ($N_\star$) at the final simulation time ($z=0.17$).
}
\label{fig:alphafig}
\end{figure*}

Fig.~\ref{fig:alphafig} shows the evolution of the \afet-\feh plane for stars residing at galactocentric radii $<20\kpc$ and $<3\kpc$ from the midplane at the indicated simulation times. The majority of stars in the high-\afe sequence develops during the first $\sim 3\Gyr$ of cosmic evolution (top row). At $z=2$ (top right panel), a dichotomy in \afe has formed for \feh$\gtrsim 0$, triggered by low metallicity gas accretion onto the then compact ($<4\kpc$) metal-rich disc. 

The formation of the outer, misaligned, metal-poor gas disc (see Section \ref{sect:whathappened} and \citetalias{Renaud2020b}) is evident in the middle row of panels ($z=1.5-1.0$), where stars can be seen to form in the low-\afe sequence at low metallicities (\feh$\sim -0.7$). At $z=1.25$, a gap exists between this metal-poor population and the rest of the low-\afe sequence; a consequence of the ISM in the inner and outer discs' being well separated spatially. Subsequent gas accretion and ISM mixing between the aligning inner and outer discs allow for a coherent low-\afe sequence to develop at $z<1$. 

We note that this is a disc-wide analysis, meant to demonstrate how distinct sequences in \afet-\feh form. The current day detectability of stellar populations of different chemical abundances depends on location in the galaxy and how efficiently stars radially migrate, as demonstrated in the main text.

\end{document}